\UseRawInputEncoding
\documentclass[prd,eqsecnum,twocolumn,nofootinbib,superscriptaddress]{revtex4-1}
\usepackage{amsfonts}
\usepackage{amsmath}
\usepackage{bm}
\usepackage{tensor}
\usepackage{ulem}
\usepackage{graphicx}
\graphicspath{ {figures/} }

\begin{document}

\title{Relativistic three-body effects in hierarchical triples}

\author{Halston Lim} 
\affiliation{Department of Physics and MIT Kavli Institute, Massachusetts Institute of Technology, Cambridge, MA 02139, USA}

\author{Carl L.~Rodriguez} 
\affiliation{Harvard Institute for Theory and Computation, 60 Garden St, Cambridge, MA 02138, USA}

\begin{abstract}
The hierarchical three-body problem has many applications in relativistic astrophysics and can play an important role in the formation of the binary black hole mergers detected by LIGO/Virgo. However, many studies have only included relativistic corrections responsible for the precession of the pericenter of the inner and outer binaries neglecting relativistic interactions between the three bodies. We revisit this problem and develop a fully consistent derivation of the secular three-body problem to first post-Newtonian order. We start with the Einstein-Infeld-Hoffman equations for a three-body system and expand the accelerations as a power series in the ratio of the semimajor axes of the inner ($a_1$) and outer ($a_2$) binary. We then perform a post-Keplerian, two-parameter expansion of the single-orbit-averaged Lagrange planetary equations in $\delta = v^2/c^2$ and $\epsilon = a_1/a_2$ using the method of multiple scales. Using this method, we derive previously indentified secular effects at $\delta \epsilon^{5/2}$ order that arise directly from the equations of motion. We also calculate new secular effects through $\delta \epsilon^4$ order that can lead to eccentricity growth over many Lidov-Kozai cycles when the tertiary is much more massive than the inner binary. In such cases, inclusion of these effects can substantially alter the evolution of three-body systems as compared to an analysis in which they are neglected. Careful analysis of post-Newtonian three-body effects will be important to understand the formation and properties of coalescing binaries that form via three-body dynamical processes.
\end{abstract}

\maketitle

\section{Introduction} \label{sec:intro}
The hierarchical three-body problem, in which a binary is orbited by a distant third companion, has wide applications in astrophysics. Triple systems can explain phenomena over a wide range of scales from asteroids to supermassive black holes (SMBHs) \cite{Wang2015INFLUENCEPLANETS,Ngo2015FRIENDSCOMPANIONS,Knutson2014FRIENDSPLANETS,Tokovinin1997,Tokovinin2014FROMSUN,Tokovinin2014FROMDWARFS,Leigh2013TheClusters,Grindlay1988Discovery1915-05,Thorsett1999TheM4,Antonini2016BLACKCLUSTERS,Kulkarni2012FormationRedshifts,Antonini2016MERGINGDETECTIONS,Deane2014ASystem}. A key characteristic of hierarchical triples is the exchange of angular momentum between the inner and outer orbits, which can lead to large inclination and eccentricity oscillations known in the literature as the Lidov-Kozai (LK) resonance \cite{Lidov1962TheBodies,Kozai1962SecularEccentricity}. Dynamical models of triples undergoing LK resonant excitations have complemented observations and informed theories about the formation and evolution of these systems, especially in the context of exoplanets and compact object mergers \cite{Naoz2012OnBinaries,Wu2007HotSystems,Thompson2011AcceleratingExotica,Antonini2012SECULAREXOTICA}.

Traditionally, the LK effect is calculated by expanding the orbit-averaged, three-body Newtonian equations of motion as a power series in $\epsilon = a_1/a_2$, where $a_1$ and $a_2$ are the semimajor axes of the inner and outer binaries, respectively. Perturbations that accumulate over each orbit (unlike periodic average-free perturbations) are referred to as ``secular" perturbations. The leading secular effect, the Newtonian-quadrupole or ``quadrupole" for short, arises at order $\epsilon^3$ beyond Keplerian forces which scale as $r^{-2}$. These quadrupole terms facilitate the exchange of orbital angular momentum which induces oscillations in the eccentricity and inclination. 

Higher-order perturbations can change the nature of the LK effect. The addition of $\epsilon^4$ (octupole) order perturbations can cause orbital flips \cite{Lithwick2011TheParticle,Naoz2011HotInteractions,Li2014EccentricitySystems}, extremely large eccentricities \cite{Ford2000,Naoz2013a, Teyssandier2013ExtremePlanets}, and chaotic evolution \cite{Lithwick2011TheParticle,Li2014ChaosMechanism}. These behaviors persist through $\epsilon^5$ (hexadecapole) order \cite{Will2017}.

The implications of two-body relativistic effects in LK triples have been thoroughly studied. In a post-Newtonian expansion of the two-body equations of motion, the leading relativistic effect induces the precession of pericenter and appears at order $\delta=v^2/c^2$ (``1pN" order) beyond Keplerian forces, where $v$ is the velocity of the inner binary. If the 1pN precession timescale of the inner binary is much shorter than the quadrupole timescale, eccentricity growth is suppressed \cite{Ford2000,Fabrycky2007ShrinkingFriction,Naoz2013}. Alternately, if the 1pN precession timescale is comparable to the quadrupole and octupole timescales, eccentricity growth is heightened \cite{Ford2000,Naoz2013}. Dissipative terms appearing at order $\delta^{5/2}$ (``2.5pN" order) cause the orbit to shrink due to gravitational radiation. Eccentricity peaks induced by the LK effect can drastically increase the efficiency of gravitational radiation, driving the inner binary to merge much faster than if the binary were circular \cite{Blaes2002, Hoffman2007DynamicsGalaxies,Antognini2014RapidTriples}. This has exciting implications for compact-object binaries with third companions as potentially eccentric gravitational-wave (GW) sources for LIGO and LISA \cite{Antonini2016MERGINGDETECTIONS, Rodriguez2018}. The outer binary's 1pN precession appears at order $\delta_2 = V^2 / c^2$ beyond Keplerian forces (in the outer binary), where $V$ is the velocity of the outer binary. While many studies have included the outer 1pN precession, it does not typically have a strong effect \cite{Naoz2013,Naoz2016}.

\begin{figure}
\centering
\includegraphics[width = .5\textwidth]{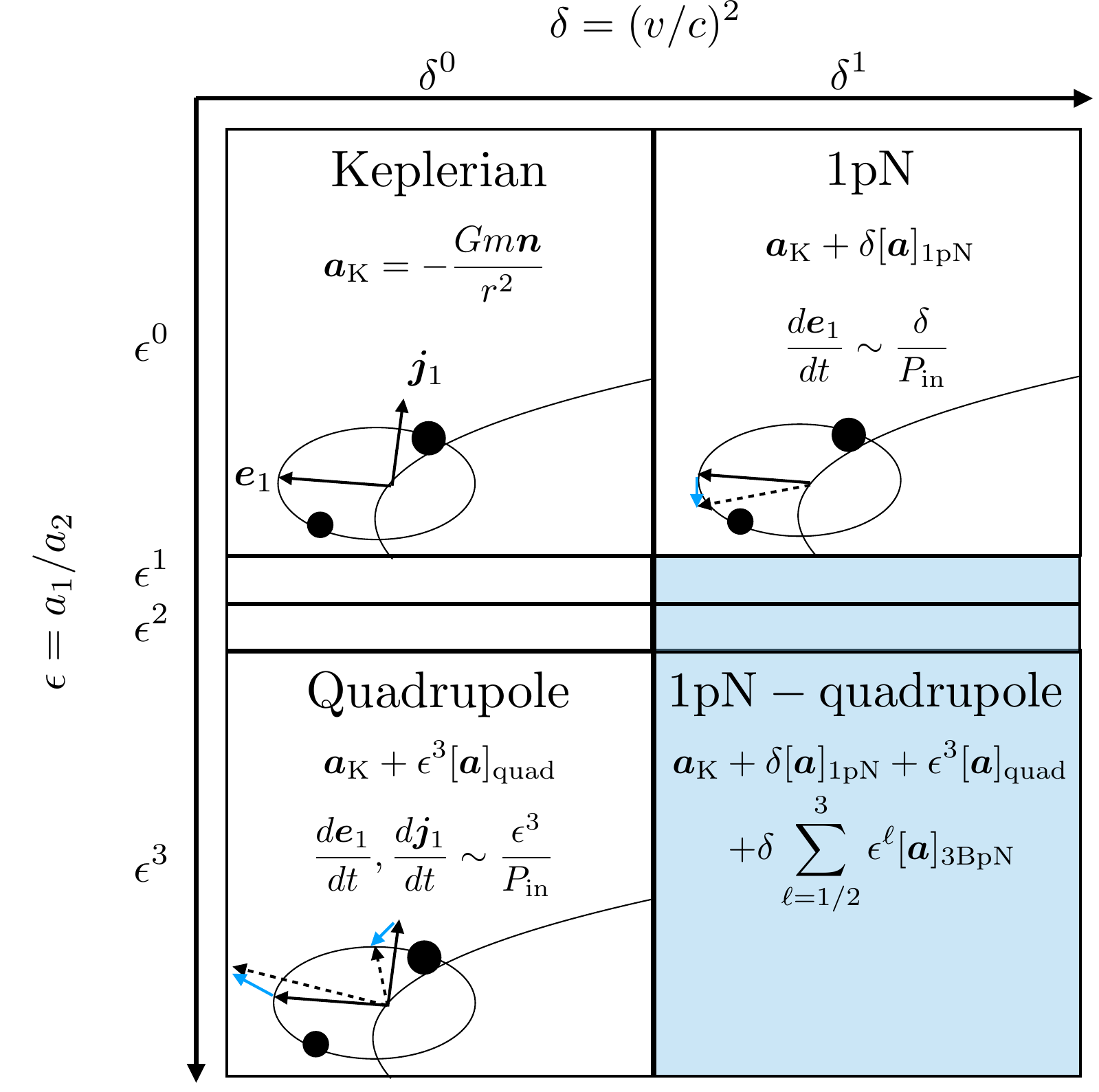}
\caption{Secular effects on the inner binary in a post-Keplerian, two-parameter expansion in the post-Newtonian parameter $\delta = (v/c)^2$ and ratio of semimajor axes $\epsilon=a_1/a_2$. At zeroth order, both the inner and the outer orbits remain fixed; the orbital vectors $\bm{e}_1 = e_1 \bm{n}$ (pointing toward the binary's pericenter) and $\bm{j}_1=\sqrt{1-e_1^2}\bm{h}$ (parallel to the orbital angular momentum) do not change. The leading three-body effects induce the LK resonance which induces perturbations $\Delta \bm{e}_1,\Delta \bm{j}_1 \propto \epsilon^3$ over a single orbit $P^{\rm K}_{\rm in} = 2\pi/\sqrt{G m/a_1^3}$. The 1pN effects on the inner binary appearing at $\delta$ order induce a precession of the pericenter. In this study, we investigate three-body 1pN effects (3BpN) due to accelerations through $\delta\epsilon^4$ ``1pN-octupole" order (which includes the shaded region). Certain 3BpN effects are distinct, such as the de Sitter precession, which causes $\bm{e}_1$ and $\bm{j}_1$ to precess about the outer orbit's angular momentum. Other 3BpN effects appear as corrections to the quadrupole and 1pN terms.}
\label{fig:expansion_table}
\end{figure}

In comparison, little is known about relativistic three-body effects or how they may alter eccentricity growth in LK triples. Even though the three-body 1pN (3BpN) terms are required for a self-consistent 1pN secular evolution, they are generally not included in the majority of analyses of hierarchical triples. The 3BpN effects can be derived with a post-Keplerian, two-parameter expansion which we illustrate in Fig.~\ref{fig:expansion_table}. We differentiate two-body 1pN (2BpN) effects from three-body 1pN (3BpN) effects as follows:
\begin{itemize}
    \item 2BpN refers to the 1pN pericenter precession effects on both the inner and the outer binaries. We consider the outer 1pN precession as a two-body effect as this effect does not depend on the inner binary separation and would happen identically if the inner binary were replaced by a single body of equivalent mass.
    \item 3BpN refers to all other 1pN effects not including the 1pN pericenter precessions.
\end{itemize}
The 3BpN terms are referred to as the ``interaction terms" by Ref.~\cite{Naoz2013} or ``cross terms" by Ref.~\cite{Will2014}.

One approach to study the 3BpN cross terms is to directly integrate the complete three-body pN equations, as done in Refs.~\cite{Galaviz2011CharacterizationHoles,Galaviz2011StabilityConfigurations,Bonetti2016Post-NewtonianTests,Lousto2008Three-bodyApproximations}. While these numerical solutions are exact, much work is required to gather physical insight. Approaches involving analytic expressions from perturbative calculations can play an important role in interpreting the output of $N$-body codes and understanding underlying physics. Such a synergy is common in the existing literature on secular effects in LK triples (e.g. the case of orbital flips in hot Jupiter systems \cite{Naoz2013,Naoz2011HotInteractions}). Furthermore, such integrations are typically far more time consuming than an integration of the secular equations, prohibiting a broad exploration of the parameter space. 

To our knowledge, only four existing studies \cite{Naoz2013,Will2014,Will2018,Liu2019BinaryTriples} investigate 3BpN cross terms on the inner binary with an orbit-averaged, perturbative approach. Although results from these studies suggest that specific 3BpN terms can significantly affect the evolution of the inner binary, they either only consider a subset of the relevant 3BpN terms \cite{Naoz2013,Liu2019BinaryTriples} or derive them restricting the outer orbit to be constant \cite{Will2014,Will2018}.

In this study, we derive the general case for arbitrary masses and orbital parameters, using a set of equations averaged over the inner orbit. We then focus on the specific case where the outer companion is much larger than the inner binary (e.g. a binary BH around a supermassive BH) and identify the specific cross terms that can influence the dynamical evolution of the inner binary. In much of the parameter space for secular hierarchical triples, 3BpN effects are subdominant to 2BpN effects and do not alter the evolution of the triple. 

In certain regions of parameter space, the magnitude of 3BpN terms can approach that of 2BpN terms and substantially change the evolution of the inner binary. The 3BpN effects coherently modulate the amplitude of quadrupole LK oscillations which can lead to a greater range in eccentricity. In systems with initially moderate inclinations, the 3BpN terms can interact with the octupole terms and cause even larger eccentricity growth and significantly reduce merger times.

The outline of this paper is as follows: In Sec.~\ref{sec:existing} we review the existing literature on relativistic cross terms in hierarchical triples. In Sec.~\ref{sec:methods} we present a derivation starting with the Einstein-Infeld-Hoffman equations for three bodies. We then conduct a multiple-scale analysis of the Lagrange planetary equations to compute secular effects through 1pN-octupole order. The derived 3BpN cross terms are presented in App.~\ref{app:details}. We provide a \textit{Mathematica} notebook upon request that contains a complete derivation. In Sec.~\ref{sec:dominant}, we discuss general features of the 3BpN effects and estimate where in parameter space their effects may be important. In Sec.~\ref{sec:casestudy} we present examples of systems where the quadrupole LK resonance is significantly altered by 3BpN effects. In Sec.~\ref{sec:population} we analyze a population of hierarchical triples including octupole terms and gravitational-wave emission to identify systematic 3BpN effects that impact a population of LK-driven mergers around a SMBH. 

\section{Existing studies on third-body 1pN effects}\label{sec:existing}
Most investigations on relativistic triples take into account post-Newtonian corrections due to binary motion along with Newtonian third-body interactions. In comparison, little is known about three-body relativistic interactions in triples. To date, the authors are aware of four previous studies that consider these effects in hierarchical triples. We summarize these studies below and then comment on how their results motivate our current work.

\begin{enumerate}
\item[(1)] Naoz \textit{et al.} \cite{Naoz2013} derives 3BpN cross terms using the orbit-averaged three-body 1pN Hamiltonian, which is calculated by applying two successive canonical transformations: the first transformation reexpresses the Hamiltonian in terms of action-angle variables (Delaunay orbital elements) and the second transformation removes periodic terms that depend on mean anomaly angles. With the orbit-averaged Hamiltonian, the time evolution is then determined through Hamilton's equations. However, as pointed out in Ref.~\cite{Will2014}, the presented orbit-averaged Hamiltonian approach may not take into account secular 3BpN effects generated indirectly from first-order variations in the orbital elements. As we will show, the leading-order periodic effects generate additional secular 3BpN cross terms.

\item[(2)] Will \cite{Will2014} uses the Lagrange planetary equations to calculate post-Keplerian perturbations as we do here. Reference \cite{Will2014} also discusses how lower-order periodic perturbations generate higher-order secular perturbations. However, this particular analysis does not systematically distinguish between secular and periodic variations, which complicates interpretation of the results \cite{WillNoTitle}. The 3BpN cross terms are also derived by restricting the outer binary's orbit to be constant, circular, and coplanar and only considering terms to leading order in the tertiary's mass, $m_3$.

\item[(3)] Will \cite{Will2018} revisits the 3BpN cross terms in application to Mercury's orbit around the Sun. This analysis uses a multiple-scale analysis to systematically account for periodic effects. Similar to Ref.~\cite{Will2014}, this analysis assumes the outer orbit is constant, circular, and coplanar, and considers effects up to linear order in $m_3$. With these assumptions, the 3BpN terms induce a precession [c. f.  their Eq.~(1)] over one inner orbit equal to
\begin{equation}\label{eq:crosswill}
\Delta \bar{\omega} = \frac{4 \pi G m_3 a^{3/2}}{c^2 R^{5/2}} + \frac{3 \pi}{4} \frac{G m_3 a^2}{c^2 R^3}\frac{28 + 47 e_1^2}{(1-e_1^2)^{3/2}},
\end{equation}
where $\bar{\omega}$ is the pericenter angle measured from a reference direction, $R$ is the circular radius of the outer tertiary, $a$ is the Mercury-Sun semimajor axis, and $m_3$ is the mass of the tertiary planet. In this paper we will investigate a different limit where the tertiary is more massive than the inner binary.

\item[(4)] Liu \textit{et al.} \cite{Liu2019BinaryTriples} considers additional relativistic interactions between the spins and orbital angular momenta in triple systems containing a SMBH with masses $m_1 = 30$, $m_2 = 20M_\odot$, and $m_3 \gtrsim 10^8$-$10^9 M_\odot$. For the inner and outer orbit they include the 1.5pN spin-orbit (Lens-Thirring) precessions. For point-particle effects they include, through analogy with spin effects, the de Sitter precession of the inner orbital plane. They write the frequency for this cross-term precession effect as
\begin{equation} \label{eq:crossliu}
\Omega_{\rm L_{\rm in}L_{\rm out}} =  \frac{3}{2} \frac{G^{3/2} m_3 (4m + 3m_3)}{c^2 \sqrt{M}  a_2^{5/2} (1-e_2^2)}.
\end{equation}
As we will show, this term is one of many cross-terms that arises naturally in our multiple-scale approach.
\end{enumerate}

Current discrepancies in the literature over the secular 3BpN cross terms exist (e.g. between Refs.~\cite{Will2014,Naoz2013}), in part, due to differences in how lower-order periodic perturbations are considered in generating higher-order secular perturbations. Therefore, our first aim is to outline a clear procedure that systematically accounts for periodic effects for general hierarchical triple configurations.

\section{Calculating 3BpN cross terms}\label{sec:methods}
\subsection{1pN Equations of Motion} \label{sec:1pNeom}
The Einstein-Infeld-Hoffman (EIH) equations describe the post-Newtonian gravitational dynamics of a system of pointlike masses. The equations are expressed in terms of coordinate positions $\bm{r}_i = r_i \bm{n}_i$ and velocities $\bm{v}_i$, where $i$ labels each mass. For a system of pointlike masses the accelerations are given by
\begin{widetext}
\begin{align}\label{eq:EIH}
\begin{split}
    \frac{d^2\bm{r}_i}{dt^2} = & -\sum\limits_{j\neq i}\frac{G m_j \bm{n}_{ij}}{r_{ij}^2} + \frac{1}{c^2}\Bigg\{\sum\limits_{j\neq i}\frac{G m_j \bm{n}_{ij}}{r_{ij}^2}  \bigg \lbrack 4 \frac{G m_j}{r_{ij}}  + 5\frac{G m_i}{r_{ij}}+\sum\limits_{k\neq i,j}\bigg(\frac{G m_k}{r_{jk}}+4\frac{G m_k}{r_{ik}} - \frac{G m_k r_{ij}}{2 r_{jk}^2}\bm{n}_{ij}\cdot \bm{n}_{jk}\bigg) \\
    & - v_{i}^2 + 4 \bm{v}_i \cdot \bm{v}_j - 2 v_j^2 + \frac{3}{2}(\bm{v}_{j}\cdot \bm{n}_{ij})^2\bigg\rbrack - \frac{7}{2}\sum\limits_{j\neq i}\frac{G m_i}{r_{ij}}\sum\limits_{k\neq i,j} \frac{G m_k \bm{n}_{jk}}{r_{jk}^2} + \sum\limits_{j\neq i} \frac{G m_j}{r_{ij}^2}\bm{n}_{ij}\cdot(4\bm{v}_i-3\bm{v}_j)(\bm{v}_{i}-\bm{v}_j) \Bigg\}, \\
\end{split}
\end{align}
\end{widetext}
where $\bm{n}_{ij} = \bm{n}_i-\bm{n}_j$ and $\bm{r}_i - \bm{r}_j=r_{ij}\bm{n}_{ij}$.

In a hierarchical triple, two bodies of mass $m_1$ and $m_2$ constitute an ``inner" orbit with separation $\bm{r}\equiv\bm{r}_{12}$ and center of mass $\bm{r}_0$. A tertiary body of mass $m_3$ follows an ``outer" orbit about the inner orbit's center of mass with separation $\bm{R}\equiv\bm{r}_3 - \bm{r}_0$, where $|\bm{R}| \gg |\bm{r}|$. For the inner and outer orbits, we define the velocities as $\bm{v}\equiv d\bm{r}/dt$, $\bm{V}\equiv d\bm{R}/dt$ and the separation unit vectors as $\bm{n}\equiv \bm{r}/r$, $\bm{N}\equiv \bm{R}/R$. In the center of mass frame,
\begin{equation}
    \sum m_i \bm{r}_i = m \bm{r}_0 + m_3 \bm{r}_3 = \mathcal{O}(c^{-2}),
\end{equation}
which leads to
\begin{align} \label{eq:relativecoords}
\begin{split}
    \bm{r}_1 & = \frac{m_2}{m} \bm{r} - \frac{m_3}{M} \bm{R}, \\
    \bm{r}_2 & = -\frac{m_1}{m} \bm{r} - \frac{m_3}{M} \bm{R}, \\
    \bm{r}_3 & = \frac{m}{M} \bm{R},
\end{split}
\end{align}
where $m = m_1+m_2$ is the total mass of the inner binary and $M = m+m_3$ is the total mass of the triple. Post-Newtonian corrections to the center of mass frame are not relevant at 1pN order since only differences of position vectors appear, and also velocities only appear in terms that are already 1pN order \cite{Will_2014}. In this frame, the 1pN acceleration of the inner orbit's center of mass $\bm{r}_0$ will also affect $\bm{R}$.

The EIH equations can be rewritten by grouping all post-Keplerian accelerations on the right-hand side,
\begin{align}
\frac{d^2\bm{R}}{dt^2} + \frac{GM}{R^2}{\bm{N}} & = \boldsymbol{A}, \label{eq:perturbkepler1} \\
\frac{d^2\bm{r}}{dt^2} + \frac{Gm}{r^2}{\bm{n}} & = \boldsymbol{a} \label{eq:perturbkepler2},
\end{align}
where $\bm{a}$ and $\bm{A}$ contain both relativistic and third-body terms. In the absence of post-Keplerian accelerations ($\bm{a},\bm{A}=0$), Eqs.~(\ref{eq:perturbkepler1}) and (\ref{eq:perturbkepler2}) take on their homogeneous forms resulting in Keplerian motion for each orbit.  

The post-Keplerian accelerations $\bm{a}$ and $\bm{A}$ contain terms that depend on powers of $r_{13},r_{23}$, which can be expanded as a power series in $\epsilon = r/R$. The Newtonian interactions between the inner and outer orbits first appear at $\epsilon^3$ order, conventionally referred to as quadrupole order in the literature \cite{Will2017}.

In the limit that $m \ll M$, perturbations on the outer binary due to the inner binary are small. Thus, for the outer binary, we only consider Newtonian three-body effects and the 2BpN term for the outer orbit,
\begin{align} 
\begin{split} \label{eq:Aexpand}
\bm{A} & = \bm{A}_{\rm 1pN} +\bm{A}_{\rm quad}+\bm{A}_{\rm oct}. \\
\end{split}
\end{align}
In contrast, perturbations on the inner binary due to the SMBH can be significant (e.g. see Ref.~\cite{Liu2019BinaryTriples}) so we include the 3BpN accelerations,
\begin{align} 
\begin{split} \label{eq:aexpand}
     \bm{a} & = \bm{a}_{\rm 1pN} +\bm{a}_{\rm quad}+\bm{a}_{\rm oct}+\bm{a}_{\rm 3BpN}. \\
\end{split}
\end{align}
The quadrupole accelerations scale relative to the Keplerian accelerations as
\begin{align} 
    \bm{a}_{\rm quad} & \sim \left(\frac{G m}{r^2}\right)\times \epsilon^3\left(\frac{m_3}{m}\right), \label{eq:newtquada}\\
    \bm{A}_{\rm quad} & \sim \left(\frac{G M}{R^2}\right)\times \epsilon^2. \label{eq:newtquadb}
\end{align}
whereas the 2BpN accelerations scale as
\begin{align}
    \bm{a}_{\rm 1pN} & \sim \left(\frac{G m}{r^2}\right)\times \delta, \label{eq:pNbinarya}\\
    \bm{A}_{\rm 1pN} & \sim \left(\frac{G M}{R^2}\right)\times \delta_2, \label{eq:pNbinaryb}
\end{align}
where $\delta = v^2/c^2 \sim (G m/r)/c^2$ is a parameter characterizing pN perturbtions on the inner binary and $\delta_2 = V^2/c^2 \sim (G M/R)/c^2$ is a parameter characterizing pN perturbtions on the outer binary. 

The 3BpN effects can arise directly from the equations of motion through $\bm{a}_{\rm 3BpN}$ or indirectly through the interaction of lower-order effects from $\bm{a}_{\rm quad}$ and $\bm{a}_{\rm 1pN}$. The interaction of lower-order perturbations on the outer binary ($\bm{A}_{\rm quad}$ and $\bm{A}_{\rm pN}$) will also induce 3BpN effects due to the coupling between the orbits. We express all pN corrections in terms of $\delta$, using 
\begin{equation}
    \delta_2 =  \delta \epsilon \left(\frac{M}{m}\right).
\end{equation}
Cross terms due to the interaction of $\bm{A}_{\rm 1pN}$ and $\bm{a}_{\rm quad}$ are order $\delta_2 \epsilon^3 \sim \delta \epsilon^4$. Therefore, we must expand the direct contributions from $\bm{a}_{\rm 3BpN}$ to comparable order $\delta \epsilon^4$:
\begin{equation}  \label{eq:directterms}
    \bm{a}_{\rm 3BpN} \sim \left(\frac{G m}{r^2}\right)\times \delta \epsilon^k \left(\frac{M}{m}\right)^\ell,
\end{equation}
where the powers of nonzero terms include
\begin{align} \label{eq:directtermspowers}
\begin{split}  
    (k,\ell) \in & \big\lbrace \left(4,2\right), \left(4,1\right),\left(4,0\right),\left(4,-1\right), \\
    & \left(\tfrac{7}{2},\tfrac{3}{2}\right), \left(\tfrac{7}{2},\tfrac{1}{2}\right), \left(\tfrac{7}{2},-\tfrac{1}{2}\right), \left(3,1\right), \left(3,0\right),\\
    &  \left(\tfrac{5}{2}, \tfrac{3}{2}\right), \left(\tfrac{5}{2},\tfrac{1}{2}\right), \left(\tfrac{5}{2},-\tfrac{1}{2}\right), \left(2,1\right), \left(2,0\right), \\
    &  \left(1,1\right), \left(1,0\right), \left(1,-1\right), \left(\tfrac{1}{2},\tfrac{1}{2} \right),\left(\tfrac{1}{2},-\tfrac{1}{2}\right) \big\rbrace.
\end{split}
\end{align}
Only the $k \geq 5/2$ terms generate nonzero secular effects. We verify that our expression for $\bm{a}_{\rm 3BpN}$ agrees with Ref.~\cite{Will2014} [c. f.  Eq.~(4.7b)] when $m_3 \ll m$.

\subsection{Lagrange Planetary Equations} \label{sec:planetaryequations}
Equations~(\ref{eq:perturbkepler1}) and (\ref{eq:perturbkepler2}) constitute a second-order differential equation for the positions and velocities of the two orbits. It is possible to rewrite this as a first-order differential equation for the time-dependent osculating orbital elements  $\{p_i,e_i,\iota_i,\omega_i,\Omega_i\}$ (e.g. see Ref.~\cite{Brouwer1961MethodsMechanics}), where $i=1,2$ labels the inner and outer orbit, respectively. The positions and velocities of each orbit are defined in terms of the orbital elements as
\begin{align} \label{eq:kepler}
\begin{split} 
    \bm{r} & = p_1 \bm{n}/ \lbrack 1+e_1 \cos (f) \rbrack, \\
    \bm{v} & = \sqrt{\frac{G m}{p_1}}\left \lbrace e_1 \sin (f) \bm{n} + \lbrack 1 + e_1 \cos (f) \rbrack \bm{\lambda} \right \rbrace,  \\
    \bm{R} & = p_2 \bm{N} / \lbrack 1+e_2 \cos (F) \rbrack, \\
    \bm{V} & = \sqrt{\frac{G M}{p_2}}\left \lbrace e_2 \sin (F) \bm{N} + \lbrack 1 + e_2 \cos (F) \rbrack \bm{\Lambda}\right \rbrace,
\end{split}
\end{align}
where the bases $\lbrace \bm{n},\bm{\lambda},\bm{h} \rbrace$ and $\lbrace \bm{N},\bm{\Lambda},\bm{H} \rbrace$ of the inner and outer orbits, respectively, can be defined with respect to a reference basis $\lbrace \bm{e}_{X},\bm{e}_{Y},\bm{e}_{Z} \rbrace$ as
\begin{align} \label{eq:elements}
\begin{split}
\bm{n} = & \left\lbrack\cos\Omega_1 \cos(\omega_1+f)-\cos\iota_1\sin\Omega_1 \sin(\omega_1+f)\right\rbrack \bm{e}_{X} \\
& + \left\lbrack\sin\Omega_1 \cos(\omega_1+f) + \cos\iota_1\cos\Omega_1 \sin(\omega_1+f)\right\rbrack \bm{e}_{Y}\\
& + \sin\iota_1\sin(\omega_1+f)\bm{e}_{Z}, \\
\bm{\lambda} = & d\bm{n}/df, \\
\bm{h} = & \bm{n} \times \bm{\lambda}, \\
\bm{N}  = & \left\lbrack\cos\Omega_2 \cos(\omega_2+F)-\cos\iota_2\sin\Omega_2\sin(\omega_2+F)\right\rbrack \bm{e}_{X}\\
& + \left\lbrack\sin\Omega_2 \cos(\omega_2+F) + \cos\iota_2\cos\Omega_2  \sin(\omega_2+F)\right\rbrack \bm{e}_{Y} \\
& + \sin\iota_2\sin(\omega_2+F)\bm{e}_{Z}, \\
\bm{\Lambda} = & d\bm{N}/dF, \\
\bm{H} = & \bm{N} \times \bm{\Lambda}.
\end{split}
\end{align}
The basis vector $\bm{e}_{Z}$ is conventionally chosen to align with the total angular momentum of the triple. The true anomalies $f$ and $F$ of the inner and outer orbits, respectively, track the phase of each orbit. $\omega_i$ is the argument of the pericenter, and $\Omega_i$ is the longitude of the ascending node.

The dynamical equations recast in terms of the above osculating orbital elements are referred to as the Lagrange planetary equations. For the inner binary, the planetary equations read
\begin{align}\label{eq:planetary}
\begin{split}
    \frac{dp_1}{dt} =& 2\sqrt{\frac{p_1}{G m}}r \mathcal{S}, \\
    \frac{de_1}{dt} =& \sqrt{\frac{p_1}{G m}} \left( \sin (f)\mathcal{R} + \frac{ 2 \cos (f) + e_1 + e_1\cos^2 (f)}{1 + e_1 \cos (f)} \mathcal{S} \right), \\
    \frac{d\omega_1}{dt} =& \frac{1}{e_1} \sqrt{\frac{p_1}{G m}} \bigg( -\cos (f) \mathcal{R}+\frac{2 + e_1\cos (f)}{1 + e_1 \cos (f)} \sin (f) \mathcal{S} \\
    & - e_1 \cot{\iota_1} \frac{\cos (\omega_1 + f)}{1 + e_1 \cos (f)} \mathcal{W} \bigg), \\
    \frac{d\iota_1}{dt} =& \sqrt{\frac{p_1}{G m}} \frac{\cos (\omega_1 + f)}{1 + e_1 \cos (f)} \mathcal{W}, \\
    \frac{d\Omega_1}{dt} =& \sqrt{\frac{p_1}{G m}} \frac{\sin (\omega_1 + f) \csc (\iota_1)}{1 + e_1 \cos (f)} \mathcal{W}, 
\end{split}
\end{align}
where 
\begin{align}\label{eq:planetarya}
\begin{split}
    \mathcal{R} & = \bm{a}\cdot \bm{n}, \\
    \mathcal{S} & = \bm{a}\cdot \bm{\lambda}, \\
    \mathcal{W} & = \bm{a}\cdot \bm{h}
\end{split}
\end{align}
are the vector components of the perturbation $\bm{a}$ projected onto the inner orbit's basis.

The planetary equations are supplemented by an additional sixth equation that converts between the true anomaly $f$ and time,
\begin{equation}
 \frac{d f}{dt} = \frac{\sqrt{G m p_1}}{r^2}-\frac{d\omega_1}{dt}-\frac{d\Omega_1}{dt}\cos \iota_1 , \label{eq:planetary2}
\end{equation}
where the first term on the right-hand side is the usual Keplerian expression and $-\dot{\omega}_1 - \dot{\Omega}_1 \cos\iota_1$ is a post-Keplerian correction.

The equations for the outer orbit are the same as Eqs.~(\ref{eq:planetary})--(\ref{eq:planetary2}), but with the substitutions $m\rightarrow M$, $\{f,p_1,e_1,\iota_1,\omega_1,\Omega_1\} \rightarrow \{F,p_2,e_2,\iota_2,\omega_2,\Omega_2\}$, and $(\mathcal{R},\mathcal{S},\mathcal{W})\rightarrow (\mathcal{R}_3,\mathcal{S}_3,\mathcal{W}_3)$, where $(\mathcal{R}_3,\mathcal{S}_3,\mathcal{W}_3) = (\bm{A}\cdot \bm{N},\bm{A}\cdot \bm{\Lambda},\bm{A}\cdot \bm{H})$ are the vector components of the perturbation $\bm{A}$ as projected onto the outer orbit's basis. Equations (\ref{eq:planetary}) and (\ref{eq:planetary2}) along with the outer orbit's counterpart equations are exact reformulations of Eqs.~(\ref{eq:perturbkepler1}) and (\ref{eq:perturbkepler2}).

From the planetary equations, one can see that inner binary perturbations that scale as 
\begin{equation} \label{eq:someacc}
    \bm{a} \sim \frac{G m}{r^2} \epsilon^k\delta^\ell
\end{equation}
generate orbital perturbations that scale as 
\begin{equation}\label{eq:someperturb}
    \frac{de_1}{dt} \sim \frac{1}{P^{\rm K}_{\rm in}} \epsilon^k\delta^\ell,
\end{equation} 
where $P^{\rm K}_{\rm in}$ is the Keplerian orbital period. Similarly, outer binary perturbations that scale as 
\begin{equation}\label{eq:outerscaling1}
    \bm{A} \sim \frac{G M}{R^2} \epsilon^k\delta^\ell
\end{equation}
generate orbital perturbations that scale as 
\begin{equation} \label{eq:outerscaling2}
    \frac{de_2}{dt} \sim \frac{1}{P^{\rm K}_{\rm out}} \epsilon^k\delta^\ell = \frac{1}{P^{\rm K}_{\rm in}} \left(\frac{M}{m}\right)^{1/2}  \epsilon^{k+3/2}\delta^\ell,
\end{equation}
where $P^{\rm K}_{\rm out}$ is the Keplerian expression for the orbital period.

Using first-order perturbation theory, the secular perturbations on the orbital elements are calculated by taking the orbit average of the planetary equations, with constant orbital elements on the right-hand side:
\begin{equation} \label{eq:timeaverage}
\left\langle \frac{d X_\alpha}{d t} \right\rangle_{t} \equiv \lim\limits_{T\rightarrow \infty} \frac{1}{T}\int_0^T \frac{d X_\alpha}{d t} dt.
\end{equation}
We use $\alpha = 1,2,...,10$ to label the orbital elements. We reserve the first five indices ($1 \leq \alpha \leq 5$) for the inner orbit's elements and the last five ($6 \leq \alpha \leq 10$) for the outer's. In the literature (e.g. Ref.~\cite{Will2017}), this integral is evaluated by using the double-orbit average approximation, which uses the fact that each term on the right-hand side of Eq.~(\ref{eq:planetary}) can be rewritten as a sum of products whose factors depend periodically on either $f$ or $F$ in addition to the orbital elements $X_\beta$:
\begin{equation}\label{eq:factor}
\frac{d X_\alpha}{d t} = \sum_i A_i(X_\beta,f) B_i(X_\beta,F).
\end{equation}
With this factorization, the average can be approximated assuming $P_{\rm in} \ll P_{\rm out}$. One first averages over the inner orbit and then subsequently averages over the outer orbit (while holding $X_\beta$ fixed),
\begin{align} 
\begin{split}
    & \left\langle \frac{d X_\alpha}{d t} \right \rangle_{t} \approx \sum_i \frac{1}{P_{\rm in}} \int_0^{P_{\rm in}} A_i dt \times \frac{1}{P_{\rm out}} \int_0^{P_{\rm out}} B_i dt \\
    & = \frac{1}{P_{\rm in} P_{\rm out}} \sum_i \int_0^{2 \pi} A_i\  \frac{dt}{df} df \times \int_0^{2 \pi} B_i\  \frac{dt}{dF}dF. \label{eq:doubleorbitavg}
\end{split}
\end{align}
The post-Keplerian corrections to $(df/dt)$, $(dF/dt)$, $P_{\rm in}$ and $P_{\rm out}$ appearing in Eq.~(\ref{eq:planetary2}) generate cross-term order effects and are not considered in first-order perturbation theory.

Before evaluating Eq.~(\ref{eq:doubleorbitavg}), the factors $A_i$ and $B_i$ can be simplified. By substituting Eqs.~(\ref{eq:kepler}) and (\ref{eq:elements}) into Eq.~(\ref{eq:planetary}), one can verify that $A_i$ and $B_i$ depend on the ascending nodes $\Omega_i$ only through powers of $\cos(\Delta\Omega)$ and $\sin(\Delta\Omega)$, where $\Delta\Omega = \Omega_1 - \Omega_2$. The equations greatly simplify by setting $\Delta\Omega = \pi$. The justification comes in two parts. First, one initially aligns the reference direction $\bm{e}_Z$ with the total orbital angular momentum so that $\Delta\Omega = \pi$. Also with this choice, Newtonian and 2BpN perturbations lead to $\dot{\Omega}_1 = \dot{\Omega}_2$ at all subsequent times. This simplification is different from eliminating the nodes in the Hamiltonian, which can lead to the incorrect equations of motion as discussed in Ref.~\cite{Naoz2013a}. The simplification we describe here is applied directly to the equations of motion. We adopt the node-eliminated simplified set of equations, but note that the cross term perturbations in general lead to $\dot{\Omega}_1 \neq \dot{\Omega}_2$. However, our quadrupole-order evolutions (Sec.~\ref{sec:casestudy}) result in $\Delta\Omega \approx \pi$ within 10\%, which provides a rough consistency check. Including corrections that depend on $\Delta\Omega$ is left to future work.

First-order perturbation theory is sufficient to calculate Newtonian secular efforts up to order $\epsilon^5$, or 1pN secular effects of order $\delta$. Second-order perturbation theory is required to calculate mixed-order ($\delta \epsilon^k$) secular effects that are generated from either lower-order periodic (average-free) perturbations or post-Keplerian corrections to $(df/dt)$, $(dF/dt)$, $P_{\rm in}$ and $P_{\rm out}$. We refer to these as the ``indirect" 3BpN cross terms, in contrast to secular effects that arise directly from the equations of motion. To calculate these periodic variations, one must solve for the instantaneous values of the elements and integrate the planetary equations with respect to an orbital phase. A few choices for the orbital phase include the true, eccentric, and mean anomalies. We use a placeholder $\phi$ to represent whatever angle is used to reparametrize the planetary equations, which read,
\begin{equation}\label{eq:lagrange_phi}
     Q_\alpha\biglb(X_\beta,F(\phi),f(\phi)\bigrb) \equiv \frac{d X_\alpha}{d\phi} = \frac{dX_\alpha}{dt} \frac{dt}{d\phi}.
\end{equation}
The planetary equations for the inner binary [Eq.~(\ref{eq:planetary})] can be organized as
\begin{align} \label{eq:perturball}
\begin{split}
     \dot{X}_\alpha = &  (\dot{X}_\alpha)_{\rm 1pN} + (\dot{X}_\alpha)_{\rm quad} +  (\dot{X}_\alpha)_{\rm 3BpN}, 
\end{split}
\end{align}
where each term on the right-hand side is due to plugging $\bm{a}_{\rm 1pN}$, $\bm{a}_{\rm quad}$, and $\bm{a}_{\rm 3BpN}$ into Eq.~(\ref{eq:planetary}), respectively. Because the scaling with $\delta$ and $\epsilon$ for each of these accelerations [Eqs.~(\ref{eq:newtquada}), (\ref{eq:pNbinarya}), and (\ref{eq:directterms})],
\begin{subequations} \label{eq:perturballscale}
\begin{eqnarray}
    \left(\dot{X}_\alpha\right)_{\rm 1pN} & \sim & \frac{\delta}{P^{\rm K}_{\rm in}} , \label{perturballscalea} \\
    \left(\dot{X}_\alpha\right)_{\rm quad} & \sim & \frac{\epsilon^3}{P^{\rm K}_{\rm in}} , \label{perturballscaleb} \\
    \left(\dot{X}_\alpha\right)_{\rm 3BpN} & \sim & \frac{\delta \epsilon^k}{P^{\rm K}_{\rm in}}. \label{perturballscalec}
\end{eqnarray}
\end{subequations}
For the outer binary [Eqs.~(\ref{eq:newtquadb}) and (\ref{eq:pNbinaryb})], the terms scale as
\begin{subequations} \label{eq:perturballscale2}
\begin{eqnarray}
\left(\dot{X}_\alpha\right)_{\rm 1pN} & \sim & \frac{\delta_2}{P^{\rm K}_{\rm out}} = \frac{\delta \epsilon ^{5/2}}{P^{\rm K}_{\rm in}}, \label{perturballscale2a} \\
\left(\dot{X}_\alpha\right)_{\rm quad} & \sim & \frac{\epsilon^2}{P^{\rm K}_{\rm out}} = \frac{\epsilon ^{7/2}}{P^{\rm K}_{\rm in}}.  \label{perturballscale2b}
\end{eqnarray}
\end{subequations}
We also include post-Keplerian corrections to $dt/d\phi$ \lbrack Eq.~(\ref{eq:planetary2})\rbrack,
\begin{equation}\label{eq:dtdphi}
    \frac{dt}{d\phi} = \left(\frac{dt}{d\phi}\right)_{\rm K} +  \left( \frac{dt}{d\phi} \right)_{\rm 1pN} + \left( \frac{dt}{d\phi} \right)_{\rm quad},
\end{equation}
where $(dt/d\phi)_{\rm K}$ is the Keplerian expression. Combining Eqs.~(\ref{eq:perturball}) and (\ref{eq:dtdphi}), we can write the re-parametrized planetary equations $Q_\alpha$ up to 1pN-quadrupole order as
\begin{align}\label{eq:perturbexpandphase}
\begin{split}
    Q_\alpha = & \left(\dot{X}_\alpha\right)_{\rm 1pN} \left \lbrack \left(\frac{dt}{d\phi}\right)_{\rm K} + \left( \frac{dt}{d\phi} \right)_{\rm quad} \right \rbrack \\
    & +\left(\dot{X}_\alpha\right)_{\rm quad} \left \lbrack \left(\frac{dt}{d\phi}\right)_{\rm K} + \left( \frac{dt}{d\phi} \right)_{\rm 1pN} \right \rbrack \\
    & + \left(\dot{X}_\alpha\right)_{\rm 3BpN} \left( \frac{dt}{d\phi} \right)_{\rm K},
\end{split}
\end{align} 
where the cross terms include
\begin{subequations} \label{eq:semidirectterms}
    \begin{eqnarray}
    & \left(\dot{X}_\alpha\right)_{\rm 1pN} \left( \dfrac{dt}{d\phi} \right)_{\rm quad}, \label{eq:semidirecttermsa}\\ 
    & \left(\dot{X}_\alpha\right)_{\rm quad} \left( \dfrac{dt}{d\phi} \right)_{\rm 1pN}, \label{eq:semidirecttermsb}\\
    & \left(\dot{X}_\alpha\right)_{\rm 3BpN} \left( \dfrac{dt}{d\phi} \right)_{\rm K}. \label{eq:semidirecttermsc}
    \end{eqnarray}
\end{subequations}
In addition to the above cross terms in Eq.~(\ref{eq:semidirectterms}), additional cross terms arise from lower-order periodic variations and corrections to the orbital periods $P_{\rm in}$ and $P_{\rm out}$. These additional cross terms can be calculated through a multiple-scale analysis described in Sec.~\ref{sec:multiscale}.

\subsection{Multiple-scale analysis} \label{sec:multiscale}
The method of multiple scales provides a clear procedure for how to systematically calculate higher-order secular effects due to lower-order periodic effects. We refer the reader to Ref.~\cite{Bender1978AdvancedEngineers} for a review of the method of multiple scales and Refs.~\cite{Mora2004Post-NewtonianObjects,Lincoln1990CoalescingEmission,Will2017} for applications in a post-Keplerian, two-body context. The multiple-scale method has also been applied to postadiabatic calculations in extreme-mass-ratio inspirals around Kerr black holes \cite{Hinderer2008Two-timescaleMotion,Will2017a}. 

In a multiple-scale analysis of the planetary equations with two bodies, one introduces an additional long-timescale variable, $\theta \equiv \epsilon \phi$, to artificially separate the secular and average-free parts of the orbital elements with the ansatz $X_\alpha = \tilde{X}_\alpha(\theta) + \epsilon W_\alpha\boldsymbol{(}\tilde{X}_\beta(\theta),\phi\boldsymbol{)}$, where $\tilde{X}_\alpha$ is the slowly evolving secular part and $W_\alpha$ is the average-free periodic part. $W_\alpha$ itself is expanded in a power series, $W_\alpha=\tensor*[]W{_{\alpha}^{(0)}} + \epsilon \tensor*[]W{_{\alpha}^{(1)}} + ...$, which can then be used to iteratively solve for $\tilde{X}_\alpha$ to desired order.

To calculate cross terms in a three-body context, one must consider perturbations by both relativistic effects and orbital interaction effects. Thus, we introduce \textit{two} long-timescale variables $\theta \equiv \epsilon \phi$ and $\tau \equiv \delta \phi$ such that
\begin{equation}
\frac{d}{d\phi} \equiv \frac{\partial}{\partial \phi} + \epsilon \frac{\partial}{\partial \theta} + \delta \frac{\partial}{\partial \tau}.
\end{equation}
The slow changing variables $\theta$ and $\tau$ resolve changes occurring over a quadrupole timescale and pN pericenter precession timescale, respectively. The fast changing variable $\phi$ describes changes occurring over an orbital period. Practical considerations which inform our choice of $\phi$ are discussed in Sec.~\ref{sec:orbitavg}. 

We introduce an ansatz to Eq.~(\ref{eq:lagrange_phi}) which reads
\begin{equation}\label{eq:ansatz}
X_\alpha\boldsymbol(\tilde{X}_\beta(\theta,\tau),\phi\boldsymbol) =  \tilde{X}_\alpha(\theta,\tau) + W_\alpha\boldsymbol(\tilde{X}_\beta(\theta,\tau),\phi\boldsymbol),
\end{equation}
where $\tilde{X}_\beta$ is the average (secular) part of $X_\alpha$ and $W_\alpha$ is the average-free (periodic) part of $X_\alpha$, defined as
\begin{align}
\langle A \rangle_{\phi} & \equiv \frac{1}{2 \pi} \int_0^{2\pi} A(\theta,\tau,\phi)\ d\phi, \label{eq:phiavg} \\
\mathcal{AF}(A) & \equiv A(\theta,\tau,\phi) - \langle A \rangle_{\phi},
\end{align}
with $\theta$ and $\tau$ held fixed in the integral.

We expand the average-free part
\begin{equation} \label{eq:avgfreepart}
W_\alpha(\tilde{X}_\beta,\phi) = \sum\limits_{\ell,m=0} \epsilon^{\ell} \delta^{m}\ \tensor*[]W{_\alpha^{\ell m}}(\tilde{X}_\beta,\phi),
\end{equation}
where $\langle \tensor*[]W{_\alpha^{\ell m}} \rangle_\phi = 0$. Note that $W^{00}_\alpha=0$ is chosen to enforce constant orbital elements at zeroth order. We substitute the ansatz [Eq.~(\ref{eq:ansatz})] back into the planetary equations [Eq.~(\ref{eq:lagrange_phi})] and separate the average part,
\begin{equation}\label{eq:secular_multiscale}
    \frac{d \tilde{X}_\alpha}{d\phi} = \left\langle Q_\alpha \right\rangle_\phi,
\end{equation}
from the average-free part,
\begin{align}\label{eq:avgfree_multiscale}
\begin{split}
    \sum\limits_{\ell,m=0}^\infty & \epsilon^{\ell} \delta^{m} \frac{\partial \tensor*[]W{_\alpha^{\ell m}}}{\partial \phi} 
    + \epsilon^{\ell+1} \delta^{m} \frac{\partial \tensor*[]W{_\alpha^{\ell m}}}{\partial \theta} + \epsilon^{\ell} \delta^{m+1} \frac{\partial \tensor*[]W{_\alpha^{\ell m}}}{\partial \tau} \\
    & =  \mathcal{AF}\big(Q_\alpha\big),
\end{split}
\end{align}
where the perturbations $Q_\alpha$ are written in Eq.~(\ref{eq:perturbexpandphase}) and we use
\begin{equation}
    \frac{d \tilde{X}_\alpha}{d\phi} = \epsilon \frac{\partial \tilde{X}_\alpha}{\partial\theta} + \delta \frac{\partial \tilde{X}_\alpha}{\partial\tau},
\end{equation}
in writing Eq.~(\ref{eq:secular_multiscale}). 

We also expand
\begin{equation} \label{eq:perturbexpandcr}
    Q_\alpha(\tilde{X}_\beta + W_\beta,\phi) = \sum\limits_{n=0}^\infty \frac{1}{n!} \frac{\partial^n Q^{(0)}}{\partial \tilde{X}_\beta ... \partial \tilde{X}_\gamma} W_\beta ... W_\gamma,
\end{equation}
where the periodic parts $W_\alpha$ are written in Eq.~(\ref{eq:avgfreepart}), repeated indices are summed over all ten elements, and
\begin{equation}
    Q^{(0)} \equiv Q(\tilde{X}_\beta,\phi).
\end{equation}
The periodic parts $W_\alpha$ combine with perturbations $Q_\alpha$ according to Eq.~(\ref{eq:perturbexpandcr}) and generate cross terms. 

Written above, Eqs.~(\ref{eq:secular_multiscale}), (\ref{eq:avgfree_multiscale}), and (\ref{eq:perturbexpandcr}) are the central equations which can be iteratively solved to obtain the secular evolution in terms of $\phi$ to desired order. To calculate the secular time evolution, one must use the conversion 
\begin{equation} \label{eq:timeconversion}
    \frac{d\tilde{X}_\alpha}{dt} =  \frac{d \tilde{X}_\alpha}{d\phi} \left \langle \frac{d\phi}{dt} \right \rangle_{\rm \phi},
\end{equation}
where the conversion factor $\left \langle d\phi/dt \right \rangle_{\rm \phi}$ also includes post-Keplerian corrections and combines with $d\tilde{X}_\alpha/d\phi$ to generate additional cross terms.

\subsection{Discussion on orbit averages} \label{sec:orbitavg}
Our discussion above is general as we did not specify the short-timescale variable $\phi$. To solve for the cross-term contributions in Eq.~(\ref{eq:secular_multiscale}) we must choose what phaselike variable to use. 

In principle, $\phi$ can be any phaselike variable characterizing the inner or outer orbits. In practice, it is difficult to explicitly write both $F$ and $f$ in terms of a single variable $\phi$. To address these difficulties, we choose $\phi = F$ and average the perturbations $Q_\alpha$ over the inner orbit, using the assumption $P_{\rm in} \ll P_{\rm out}$. This expresses the equations of motion in terms of $F$ only:
\begin{align}\label{eq:singleorbitavg}
\begin{split}
    Q_\alpha\left(X_\beta,f,F\right) = \frac{dX_\alpha}{dt} \frac{dt}{dF} \approx \left\langle \frac{dX_\alpha}{dt} \frac{dt}{dF} \right\rangle_{\rm in},
\end{split}
\end{align}
where the inner-orbit average is
\begin{equation} \label{eq:innerorbitavg}
        \left\langle A \right\rangle_{\rm in} = \frac{1}{P_{\rm in}} \int_0^{2\pi} A\left(X_\beta,f,F\right) \frac{dt}{df}\ df,
\end{equation}
with the inner period is defined as
\begin{equation}
    P_{\rm in} = \int_0^{2\pi} \frac{dt}{df}\ df,
\end{equation}
holding $F$ fixed. The orbit-average defined in Eq.~(\ref{eq:phiavg}) when evaluated with Eq.~(\ref{eq:singleorbitavg}) is also consistent with the usual double-orbit average encountered in the literature [Eq.~(\ref{eq:doubleorbitavg})]. 

In the inner-orbit average [Eq.~(\ref{eq:innerorbitavg})], we include post-Keplerian corrections to $(dt/df)$ which combine with $\dot{X}_\alpha$ to generate additional cross terms:
\begin{subequations} \label{eq:semidirectterms2}
    \begin{align}
    & (\dot{X}_\alpha)_{\rm 1pN} \left( \dfrac{dt}{df} \right)_{\rm quad}, \label{eq:semidirectterms2a}\\ 
    & (\dot{X}_\alpha)_{\rm quad} \left( \dfrac{dt}{df} \right)_{\rm 1pN}. \label{eq:semidirectterms2b}
    \end{align}
\end{subequations}
Cross terms also result from post-Keplerian corrections to the orbital period:
\begin{subequations} \label{eq:periodcorrection}
    \begin{align}
    & P^{\rm 1pN}_{\rm in} = \int_0^{2\pi} \left(\dfrac{dt}{df}\right)_{\rm 1pN}\ df, \label{eq:periodcorrectiona}\\ 
    & P^{\rm quad}_{\rm in} = \int_0^{2\pi} \left(\dfrac{dt}{df}\right)_{\rm quad}\ df, \label{eq:periodcorrectionb}
    \end{align}
\end{subequations}
which are order $\delta$ [Eq.~(\ref{eq:periodcorrectiona})] and $\epsilon^3$ [Eq.~(\ref{eq:periodcorrectionb})]  beyond the Keplerian period $P^{\rm K}_{\rm in}$. Collecting the post-Keplerian corrections, the perturbations $Q_\alpha$ can be written up to 1pN-quadrupole order as 
\begin{equation}
    Q_\alpha(X_\beta,F) = (Q_\alpha)_{\rm 1pN} + (Q_\alpha)_{\rm quad} + (Q_\alpha)_{\rm 3BpN},
\end{equation}
where 
\begin{align}\label{eq:semidirected}
\begin{split}
    (Q_\alpha)_{\rm quad} & = \frac{1}{P^{\rm K}_{\rm in}} \int_0^{2\pi} 
      \left(\dot{X}_\alpha\right)_{\rm quad} \left( \frac{dt}{df} \right)_{\rm K} \left( \frac{dt}{dF} \right)_{\rm K} df, \\
    (Q_\alpha)_{\rm 1pN} & = \frac{1}{P^{\rm K}_{\rm in}} \int_0^{2\pi} 
      \left(\dot{X}_\alpha\right)_{\rm 1pN} \left( \frac{dt}{df} \right)_{\rm K} \left( \frac{dt}{dF} \right)_{\rm K} df,
\end{split}
\end{align}
\begin{widetext}
\begin{align}\label{eq:allsemidirect}
\begin{split}
    \left(Q_\alpha\right)_{\rm 3BpN} = \frac{1}{P^K_{\rm in}} \int_0^{2\pi} \Bigg \lbrack & 
      \left(\dot{X}_\alpha\right)_{\rm 3BpN} \left( \frac{dt}{df} \right)_{\rm K} \left( \frac{dt}{dF} \right)_{\rm K} + \left(\dot{X}_\alpha\right)_{\rm 1pN} \left( \frac{dt}{df} \right)_{\rm quad} \left( \frac{dt}{dF} \right)_{\rm K} + \left(\dot{X}_\alpha\right)_{\rm quad} \left( \frac{dt}{df} \right)_{\rm 1pN} \left( \frac{dt}{dF} \right)_{\rm K} \\
    + & \left(\dot{X}_\alpha\right)_{\rm 1pN} \left( \frac{dt}{df} \right)_{\rm K} \left( \frac{dt}{dF} \right)_{\rm quad} + \left(\dot{X}_\alpha\right)_{\rm quad} \left( \frac{dt}{df} \right)_{\rm K} \left( \frac{dt}{dF} \right)_{\rm 1pN} \Bigg \rbrack df \\
    - \frac{P^{\rm 1pN}_{\rm in}}{\left(P^{\rm K}_{\rm in}\right)^2} \int_0^{2\pi} \Bigg \lbrack & 
      \left(\dot{X}_\alpha\right)_{\rm quad} \left( \frac{dt}{df} \right)_{\rm K} \left( \frac{dt}{dF} \right)_{\rm K} \Bigg \rbrack df - \frac{P^{\rm quad}_{\rm in}}{\left(P^{\rm K}_{\rm in}\right)^2} \int_0^{2\pi} \Bigg \lbrack
      \left(\dot{X}_\alpha\right)_{\rm 1pN} \left( \frac{dt}{df} \right)_{\rm K} \left( \frac{dt}{dF} \right)_{\rm K} \Bigg \rbrack df,
\end{split}
\end{align}
\end{widetext}
where the last two terms are from post-Keplerian corrections to $P_{\rm in}$.

The leading mixed-order secular terms in $ \left(Q_\alpha\right)_{\rm 3BpN}$ come from taking the orbit average $\langle(Q^{(0)}_\alpha)_{\rm 3BpN}\rangle_{\rm F}$. Additional cross terms arise the interaction of $(Q_\alpha)_{\rm 1pN}$ and $(Q_\alpha)_{\rm quad}$ with periodic variations $W_\alpha$ [Eq.~(\ref{eq:perturbexpandcr})].

A multiple-scale analysis of the single-orbit-averaged equations accounts for average-free perturbations periodic with $F$ but neglects those periodic with $f$. We leave an investigation of the average-free $f$-periodic variations to future work, but point out that the single-orbit-averaged equations have been shown to agree well with $N$-body integrations in the Newtonian test-particle limit ($m_2 \ll m$) \cite{Luo2016Double-averagingCorrection}. 

The leading periodic parts are $W^{01}_\alpha$ and $W^{30}_\alpha$ for the inner binary and $W^{\frac{5}{2}1}_\alpha$ and $W^{\frac{7}{2}0}_\alpha$ for the outer binary. This can be shown by combining Eqs.~(\ref{eq:perturballscale}), (\ref{eq:perturballscale2}) and (\ref{eq:singleorbitavg}), which leads to the expansions
\begin{align}
\begin{split}
   \big ( Q_{\alpha} \big )_{\rm 1pN} = \big ( Q_{\alpha}^{(0)} \big)_{\rm 1pN} & + \sum\limits_{\beta = 1}^{5} \frac{\partial \big( Q_{\alpha}^{(0)} \big)_{\rm 1pN}}{\partial \tilde{X}_\beta} W_\beta^{30} \\
    & +  \sum\limits_{\beta = 6}^{10}  \frac{\partial \big( Q_{\alpha}^{(0)} \big)_{\rm 1pN}}{\partial \tilde{X}_\beta} W_\beta^{\frac{7}{2}0}, 
\end{split}
\end{align}
\begin{align}
\begin{split}
    \big( Q_{\alpha} \big)_{\rm quad} =  ( Q_{\alpha}^{(0)} )_{\rm quad} & + \sum\limits_{\beta = 1}^{5} \frac{\partial \big( Q_{\alpha}^{(0)} )_{\rm quad}}{\partial \tilde{X}_\beta} W_\beta^{01} \\
    & + \sum\limits_{\beta = 6}^{10}  \frac{\partial \big( Q_{\alpha}^{(0)} )_{\rm quad}}{\partial \tilde{X}_\beta} W_\beta^{\frac{5}{2}1},
\end{split}
\end{align}
where the lowest-order periodic parts are,
\begin{align}
    \tensor*[]W{_\alpha^{30}} & = \int_0^{F} \mathcal{AF} \biglb( (Q_\alpha^{(0)})_{\rm quad} \bigrb) dF' + C,  \label{eq:avgfree_part1} \\
    \tensor*[]W{_\alpha^{01}} & = \int_0^{F} \mathcal{AF} \biglb( (Q_\alpha^{(0)})_{\rm 1pN}\bigrb) dF' + D, \\
    & \qquad \qquad \qquad \qquad \quad 1 \leq \alpha \leq 5, \label{eq:avgfree_part2}
\end{align}
 for the inner binary, and 
\begin{align}
    \tensor*[]W{_\alpha^{\frac{7}{2}0}} & = \int_0^{F} \mathcal{AF} \biglb( (Q_\alpha^{(0)})_{\rm quad} \bigrb) dF' + E, \label{eq:avgfree_part3} \\
    \tensor*[]W{_\alpha^{\frac{5}{2}1}} & = \int_0^{F} \mathcal{AF} \biglb( (Q_\alpha^{(0)})_{\rm 1pN} \bigrb) dF' + H, \\
    & \qquad \qquad \qquad \qquad \quad 6 \leq \alpha \leq 10,
    \label{eq:avgfree_part4}
\end{align}
for the outer binary. The integration constants $C,D,E,H$ are determined by $\langle \tensor*[]W{_\alpha^{\ell m}} \rangle_F  = 0$ and are identical to those in Eq.~(B11) in Ref.~\cite{Will2017a}.

The total 3BpN secular contribution is
\begin{widetext}
\begin{align} \label{eq:cross_multiscale}
\begin{split}
    \left(\frac{d \tilde{X}_\alpha}{dF}\right)_{\rm 3BpN} = \Bigg \langle   \sum\limits_{\beta = 1}^{5} \Bigg(\frac{\partial ( Q_{\alpha}^{(0)} )_{\rm 1pN}}{\partial \tilde{X}_\beta} W_\beta^{30} +  \frac{\partial ( Q_{\alpha}^{(0)} )_{\rm quad}}{\partial \tilde{X}_\beta} W_\beta^{01} \Bigg) + \sum\limits_{\beta = 6}^{10} \Bigg(  \frac{\partial ( Q_{\alpha}^{(0)} )_{\rm 1pN}}{\partial \tilde{X}_\beta} W_\beta^{\frac{7}{2}0} & + \frac{\partial ( Q_{\alpha}^{(0)} )_{\rm quad}}{\partial \tilde{X}_\beta} W_\beta^{\frac{5}{2}1} \Bigg) \\
    & + (Q^{(0)}_\alpha)_{\rm 3BpN}\Bigg \rangle_F.
\end{split}
\end{align}
\end{widetext}
The first two terms on the right-hand side of Eq.~(\ref{eq:cross_multiscale}) are from periodic variations of the inner orbit's elements. The third and fourth terms are from periodic variations of the outer orbit's elements. The last term is directly from the equations of motion $\bm{a}_{\rm 3BpN}$ and from corrections to $(dt/df)$, $(dt/dF)$, $P_{\rm in}$, and $P_{\rm out}$ [Eq.~(\ref{eq:allsemidirect})].

In App.~\ref{app:details} we present the cross terms as an average time derivative using Eq.~(\ref{eq:timeconversion}). With $\phi = F$, and converting between $F$ and $t$,
\begin{equation}
    \frac{d\tilde{X}_\alpha}{dt} =  \frac{d \tilde{X}_\alpha}{dF} \left \langle \frac{dF}{dt} \right \rangle_{F} = \frac{d \tilde{X}_\alpha}{dF} \frac{ 2 \pi}{P_{\rm out}}.
\end{equation}
The outer orbital period $P_{\rm out}$can be calculated with the single-orbit average approximation [Eq.~(\ref{eq:innerorbitavg})] as
\begin{align}
\begin{split}
    P_{\rm out} & \equiv 2 \pi  \left \langle \frac{dt}{dF} \right \rangle_F = \int\limits_0^{2\pi} \left(\frac{dt}{dF}\right)\ dF \approx \int\limits_0^{2\pi} \left \langle \frac{dt}{dF} \right \rangle_{\rm in}\ dF.
\end{split}
\end{align}
We take into account leading-order corrections to $P_{\rm out}$ from the standard corrections to $(dt/dF)$ [Eq.~(\ref{eq:planetary2})] and also periodic variations in $(dt/dF)$.
\begin{figure*}
\centering
\includegraphics[width = 1\textwidth]{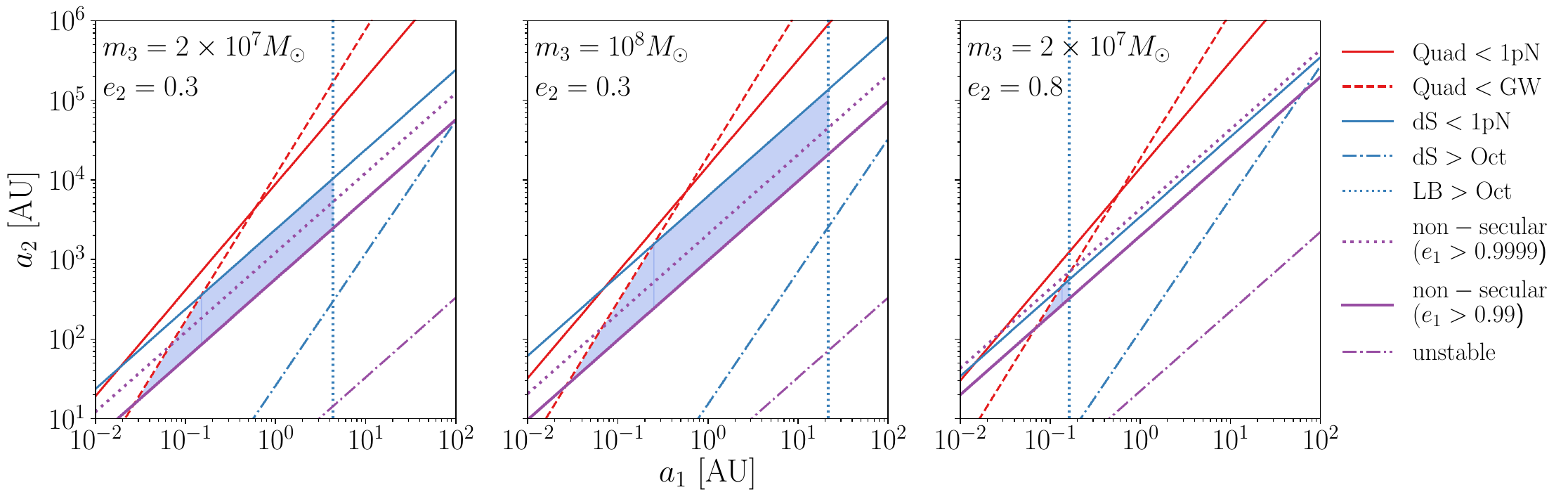}
\caption{Estimated regions in $(a_1,a_2)$ in parameter space where 3BpN cross terms are expected to be significant in hierarchical triples with a SMBH. Each line is found by equating the timescales (or scaling magnitudes $\mathcal{X}_{\ell m}$) of various effects. The descriptions corresponding to each line describe the relative timescale of two secular effects in the region to the right of the line. For example, the solid red line (Quad $<$ 1pN) marks where the quadrupole (LK) and inner 1pN precession effects operate on the same timescale; to the right of this line (for larger $a_1$, smaller $a_2$) LK effects dominate over 1pN precession in the inner binary (using criterion from Ref.~\cite{Antonini2018PrecessionalSpin}). We shade the region of secular parameter space where the de Sitter (dS) cross terms are at least comparable to both 1pN precession and octupole (Oct) effects, and also where the librating cross terms (LB) are at least comparable to octupole effects. }
\label{fig:timescales}
\end{figure*}

\section{Effects of Third-body 1pN Cross Terms due to a SMBH} \label{sec:results}
\subsection{Dominant cross terms around a SMBH} \label{sec:dominant}

For completeness, we keep cross terms of all powers in $(m/M)$ in the derivation in Sec.~\ref{sec:methods}, but we work on the assumption that the inner binary's mass is small relative to the total mass [Eqs.~(\ref{eq:Aexpand}) and (\ref{eq:aexpand})]. In this Section, we closely examine the dominant cross-term effects when $m \ll M$ and locate regions in parameter space where their effects become significant in triples undergoing strong LK oscillations.

We are in particular interested in how the dominant three-body 1pN (3BpN) cross terms interact with other secular effects, including the two-body 1pN (2BpN), quadrupole, and octupole terms. The conventional picture is that 1pN pericenter precession in the inner binary will quench eccentricity growth if the timescale for precession is much shorter than that of quadrupole (LK) effects \cite{Blaes2002,Merritt2013DynamicsNuclei}.

However, in some cases, inner 1pN effects can instead stimulate eccentricity growth. Refs.~\cite{Ford2000} and \cite{Naoz2013} demonstrate heightened resonantlike eccentricity excitation if the inner orbit's 1pN precession timescale is comparable to the Newtonian (quadrupole and octupole) timescales. Given this resonantlike behavior between the inner 1pN and Newtonian terms, it may be unsurprising if the 3BpN cross terms also lead to resonantlike behavior when their effective timescale approaches that of inner 1pN or Newtonian effects. 

The mixed-order ($\delta \epsilon^k$) cross terms are higher order than the inner 1pN ($\delta$) terms. But as $q=M/m$ increases, so does the relative strength of cross terms which scale with positive powers of $q$. We consider the contribution of the cross terms relative to the inner 1pN precession effect,
\begin{equation} \label{eq:perip}
\left(\frac{d\omega_1}{dt}\right)_{\rm 1pN} = \frac{G^{3/2}m^{3/2}}{c^2 a_1^{5/2} (1-e_1^2)} \sim \frac{\delta}{P_{\rm in}},
\end{equation}
such that the total contribution from all cross terms reads
\begin{align}\begin{split}\label{eq:scaling}
\left( \frac{dX_\alpha}{dt} \right)_{\rm 3BpN} & = \frac{\delta}{P_{\rm in}} \sum\limits_{\ell,m}  f^\alpha_{\ell m}\ \mathcal{X}_{\ell m},
\end{split}\end{align}
where $f_\alpha^{\ell m}$ contains numerical factors of order unity and factors including $e_j,\omega_j,\iota_j$. The scaling magnitude,
\begin{equation}
    {\mathcal{X}}_{\ell m} = \left(\frac{M}{m}\right)^\ell \left(\frac{a_1}{a_2}\right)^m,
\end{equation}
can be used as an estimate for which cross terms will be dominant or subdominant given an initial set of triple parameters. Since the semilatus rectum has dimensions of length, we compensate by defining $f^{p_1}_{\ell m}$ with an additional factor of $p_1$ so that a given perturbation leads to the same scaling factor ${\mathcal{X}}_{\ell m}$ across all elements. The inner 1pN precession term [Eq.~(\ref{eq:perip})] has a scaling magnitude of $\mathcal{X}_{00} = 1$.

We compare the cross-term scaling magnitudes for a generic hierarchical triple system with  $m_1=m_2=25 M_\odot, m_3 = 4\times 10^6 M_\odot$, and initial semimajor axes $a_1 = 1\ {\rm AU}$ and $a_2 = 2000\ {\rm AU}$. For a wide portion of parameter space, when $m \ll M$, the dominant 3BpN effect on the inner binary is the geodetic (de Sitter-like) precession of the inner orbit's vectors, $\bm{e}_1$ and $\bm{j}_1$, as they are parallel transported around the SMBH. This de Sitter cross term, which comes directly from the EIH equations ($\bm{a}_{\rm 3BpN}$), induces the orbital element $\bar{\omega}_1 \equiv \omega_1 + \Omega_1\cos\iota_1$ to precess at the rate
\begin{equation}\label{eq:ct0}
    \Big(\frac{d \bar{\omega}_1}{dt}\Big)_{\rm 3BpN} = \frac{G^{3/2}(4 m + 3 m_3)m_3}{2 M^{1/2} c^2 a_2^{5/2} (1-e_2^2)} \cos\iota,
\end{equation}
and has a scaling magnitude $\mathcal{X}_{\frac{3}{2}\frac{5}{2}} = 0.13$, ignoring smaller corrections proportional to $\mathcal{X}_{\frac{1}{2}\frac{5}{2}} = 10^{-5} \mathcal{X}_{\frac{3}{2}\frac{5}{2}}$. 

For the same initial parameters, the second dominant cross-term effect perturbs the pericenter at a rate
\begin{align} \label{eq:ctn1}
\begin{split}
    & \Big(\frac{d \omega_1}{dt}\Big)_{\rm 3BpN} = \frac{15 G^{3/2}m_3 m}{4 M^{1/2} a_1 a_2^{3/2} c^2}\frac{e_1^2 (1 + \ell_2-2\ell_2^2)}{\ell_1^2 (1+\ell_2)} \\
    & \times \Big(\cos\iota \cos2\omega_1 \cos2\omega_2 + \frac{1+\cos^2\iota}{2}\sin2\omega_1\sin2\omega_2 \Big),
\end{split}
\end{align}
where $\ell_1 = \sqrt{1-e_1^2}$ and $\ell_2 = \sqrt{1-e_2^2}$, and has a scaling magnitude $\mathcal{X}_{\frac{1}{2}\frac{3}{2}} = 3\times 10^{-3}$. In isolation, the cross-term effect in Eq.~(\ref{eq:ctn1}) will lead to bounded oscillations in $\omega_1$. For this reason we refer to Eq.~(\ref{eq:ctn1}) as the ``libration" cross term. The term arises from the interaction of the inner 1pN precession with outer quadrupole effects. Note that if the 1pN-binary precession [Eq.~(\ref{eq:perip})] is dominant, the libration cross term will average out.

The third dominant cross terms arise from the interaction of inner quadrupole effects with the outer 1pN precession. Unlike the previous two cross terms, these perturbations affect all inner orbital elements, and are presented in App.~\ref{app:details}; as an example, we write the perturbation on eccentricity below:
\begin{align} \label{eq:ct1p5}
\begin{split}
    & \Big(\frac{de_1}{dt}\Big)_{\rm 3BpN} =
    \frac{15 G^{3/2} M^2 a_1^{3/2}}{32 c^2 m^{1/2} a_2^4} \frac{e_1\ell_1}{\ell_2^7} \bigg( e_2^2 \ell_2^2 \big\lbrack (3 + \cos2\iota) \cos2\omega_2 \\
    & \times \sin2\omega_1 - 4 \cos\iota \cos2\omega_1 \sin2\omega_2 \big\rbrack - g_{e_2} \sin^2\iota \sin2\omega_1 \bigg)
\end{split}
\end{align}
where
\begin{equation} \label{eq:ct1p5b}
    g_{e_2} = 6 (8 + 3 e_2^2 + 4 e_2^4).
\end{equation}
These effects have scaling magnitude $\mathcal{X}_{24} = 4\times10^{-4}$, and are the leading relativistic corrections to the quadrupole effect. Therefore, we call terms that scale as $\mathcal{X}_{24}$ as ``relativistic-LK" cross terms. The scaling $\mathcal{X}_{24}$ suggests that the relativistic-LK cross terms will surpass the libration cross terms in magnitude when $q>\epsilon^{-5/3}$. Other cross terms besides those written in Eqs.~(\ref{eq:ct0})--(\ref{eq:ct1p5}) are negligible with scaling magnitudes $\mathcal{X}_{31},\mathcal{X}_{20} \leq 10^{-5}$. 

Although the scaling magnitudes $\mathcal{X}_{\ell m}$ quoted above are specific to a system with initial parameters $(a_1,a_2,m,M) = (1{\rm AU}, 10^4{\rm AU}, 50M_\odot, 4\times 10^6 M_\odot)$,  the general conclusion is the same in much of parameter space: the de Sitter, libration, and relativistic-LK cross terms [Eqs.~(\ref{eq:ct0})--(\ref{eq:ct1p5})] represent the dominant relativistic three-body secular effects. For the remainder of the paper, we focus on the effect of these three dominant cross terms and neglect other subdominant cross terms.

Inspired by recent direct detections made by LIGO, we choose $m_1 = 30 M_\odot$ and $m_2 = 20 M_\odot$. As the mass ratio $q$ increases, so does the region of ($a_1,a_2$) parameter space where cross terms are expected to be significant. When $q \gtrsim 10^7$, the resolution required to resolve quadrupole effects becomes computationally burdensome, as the quadrupole timescale goes as $\sqrt{m}/M$ \cite{Naoz2013}.

Given the masses and initial eccentricities, we can identify regions in parameter space where the cross terms are significant by comparing timescales for various effects (Fig.~\ref{fig:timescales}). Our primary interest lies in triples where the LK effects may lead to eccentricity growth, so we demand that the inner 1pN precession not squash LK effects (c.f. Eq.~(10) in Ref.~\cite{Antonini2018PrecessionalSpin}). Another constraint we impose is that the GW timescale is longer than the LK timescale (c.f. Eq.~(31) in Ref.~\cite{Rodriguez2018}). We must also stay in the region of parameter space where the secular approximation is valid. We use the criterion from Ref.~\cite{Antonini2014} and restrict our initial parameters assuming the maximum eccentricity achieved is $e_1 = 0.99$. This limit is somewhat arbitrary, since we also verify the secular criterion for each evolution \textit{a posteriori}. Finally, we estimate where the de Sitter precession rate exceeds the inner 1pN prececssion rate, and where the librating cross term exceeds the octupole terms. We set $ m_3 = 2\times10^7\ M_\odot$ and $e_2 = 0.8$, and we leave a wider exploration of parameter space and larger $m_3$ to future work.

\subsection{Case study} \label{sec:casestudy}
In this section, we discuss two examples of resonantlike behaviors induced by the 3BpN terms. We demonstrate the effect of these behaviors by comparing evolutions: one with and without 3BpN cross terms. We restrict our attention to the three dominant cross-term effects discussed in Sec.~\ref{sec:dominant} and initially neglect octupole effects and GW dissipation. Later in Sec.~\ref{sec:population} we discuss the 3BpN effects conjunction with octupole effects and GW dissipation.
\begin{figure}
\centering
\includegraphics[width = .5\textwidth]{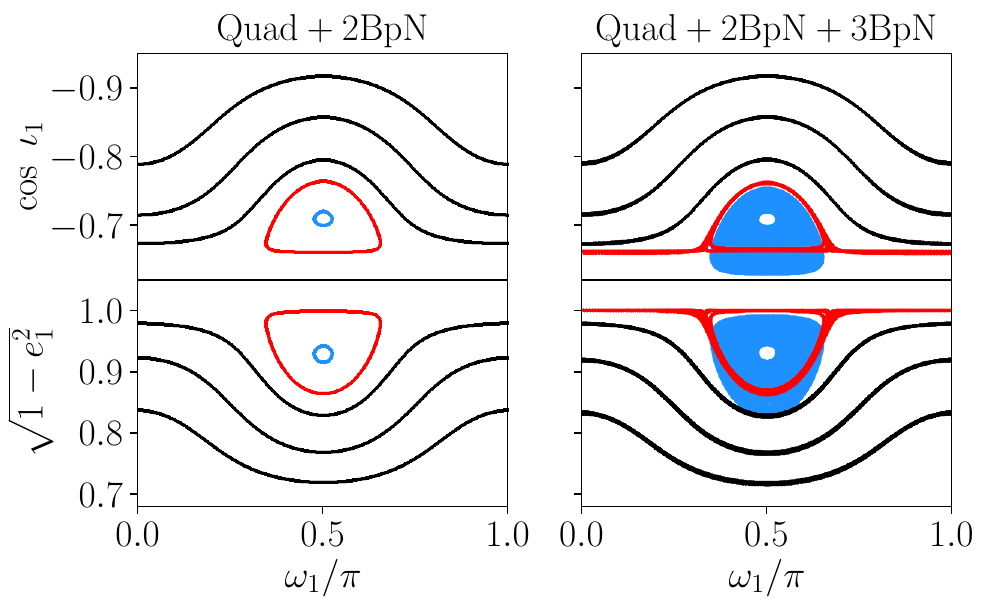}
\caption{Three-body 1pN (3BpN) effects for a librating system, including quadrupole (Quad), inner and outer 1pN precessions (2BpN), and 3BpN effects. We plot trajectories in $(\iota_1,\omega_1)$ phase space (top) and $(e_1,\omega_1)$  phase space (bottom) for a triple with $(m_1,m_2,m_3)=(30M_\odot,20M_\odot,2\times10^7M_\odot)$, $e_2 = 0.8$, and $(a_1,\ a_2) = (0.10\ {\rm AU}, 209.84\ {\rm AU})$. Each trajectory is initialized with $\omega_1 = 90^\circ$, $\omega_2 = 282.27^\circ$, $\Omega_1 = 192.5^\circ$, and $\Omega_2 =  12.5^\circ$ but with different initial $e_1$ and $\iota_1$ such that $\ell_{z} = \sqrt{1-e_1} \cos\iota_1 = -0.6593$. In the limit that $m/M \ll 1$ and $a_1/a_2 \ll 1$, when only considering 2BpN and quadrupole effects, the z-component of the angular momentum of the inner binary $\ell_z$ [Eq.~(\ref{eq:lzdef})] is nearly constant (left). As a result, all trajectories are closed and either exhibit circulation or libration. 3BpN effects lead to thickening of the phase space trajectories. For librating trajectories inside the separatrix, 3BpN effects can significantly modulate the amplitude of LK cycles, filling nearby regions of phase space (blue). The time evolution of the blue trajectory is plotted in Fig.~\ref{fig:timespace_librating}. Trajectories near the separatrix switch between librating and circulating (red). }
\label{fig:phasespace_librating}
\end{figure}
\begin{figure}
\centering
\includegraphics[width = .45\textwidth]{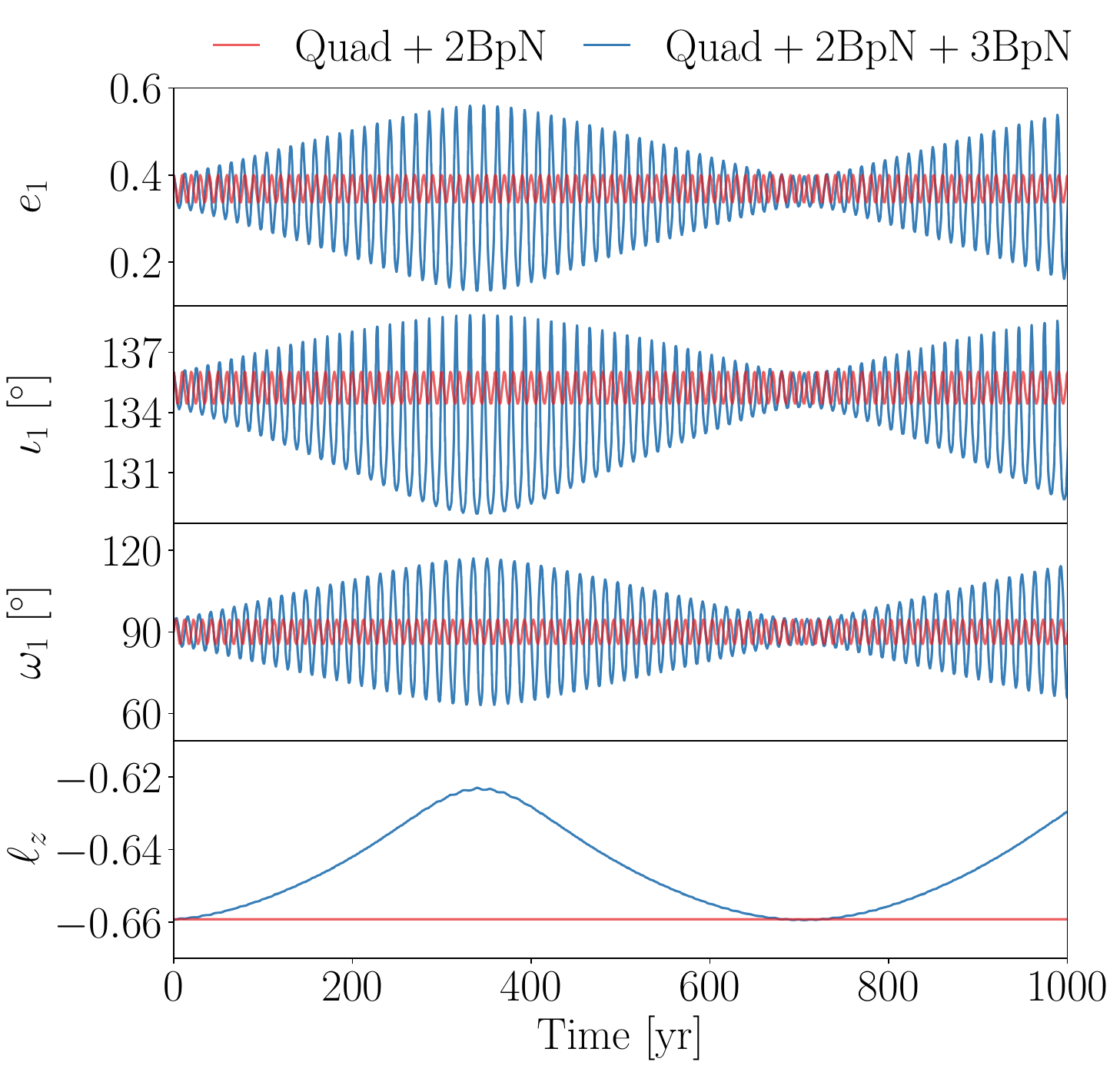}
\includegraphics[width = .48\textwidth]{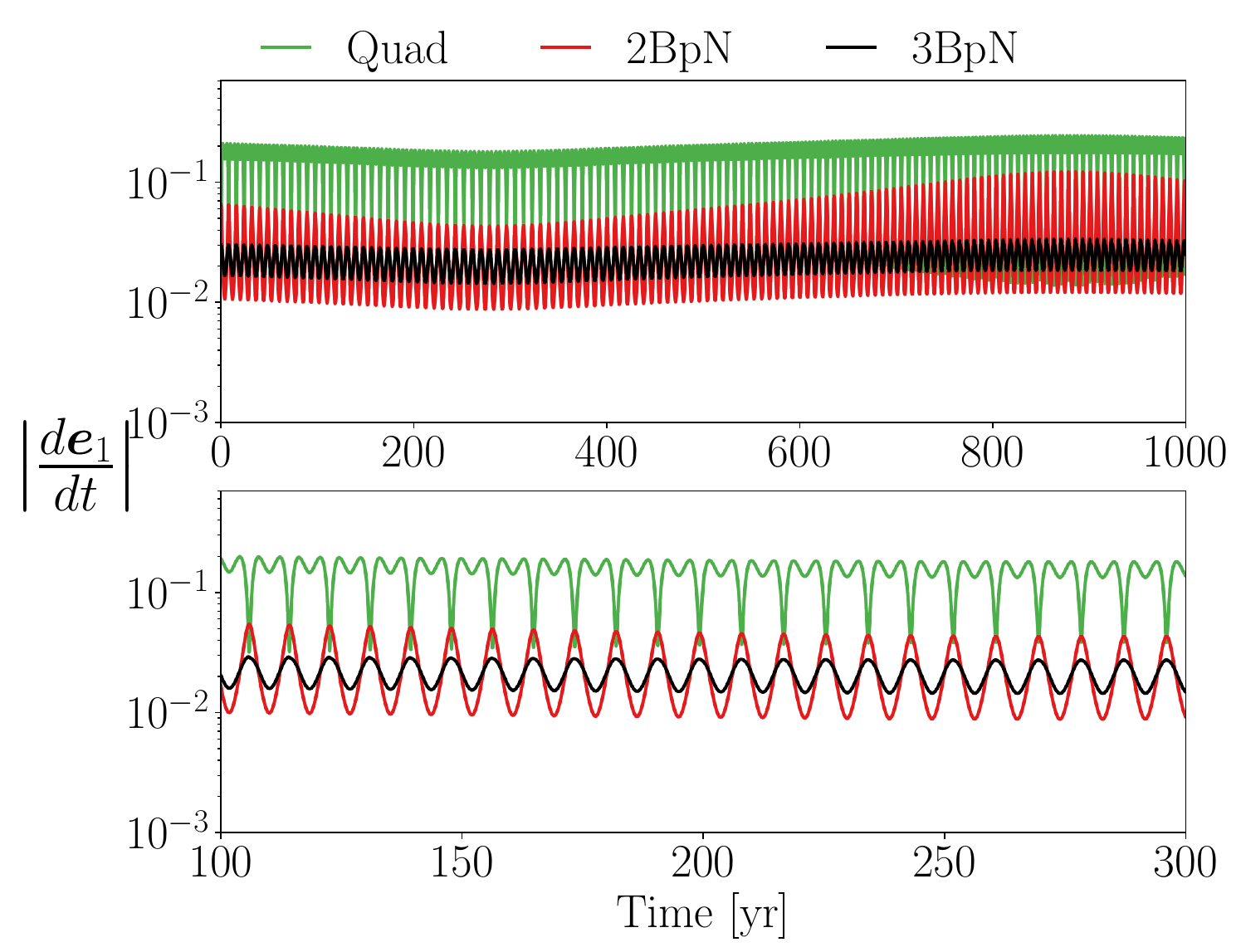}
\caption{Evolution of a triple system exhibiting the 3BpN librating resonance. We plot the time evolution (top) including quadrupole and 2BpN effects (red), and the time evolution including quadrupole, 2BpN, and 3BpN effects (blue) on the inner binary. With 3BpN effects, the angular momentum component $\ell_z$ oscillates about its initial value. This induces modulations in the amplitude of LK oscillations in $e_1$, $\iota_1$,  and $\omega_1$, which reach a maximum when $\ell_z$ is at a maximum (e.g. near $t=350\ {\rm yr}$). We also show the instantaneous magnitudes of quadrupole, 2BpN and all 3BpN perturbations on $\dot{\bm{e}}_1$ [Eq.~(\ref{eq:edotdef})] in units of ${\rm yr}^{-1}$ (bottom).} 
\label{fig:timespace_librating}
\end{figure}
\begin{figure}
\centering
\includegraphics[width = .5\textwidth]{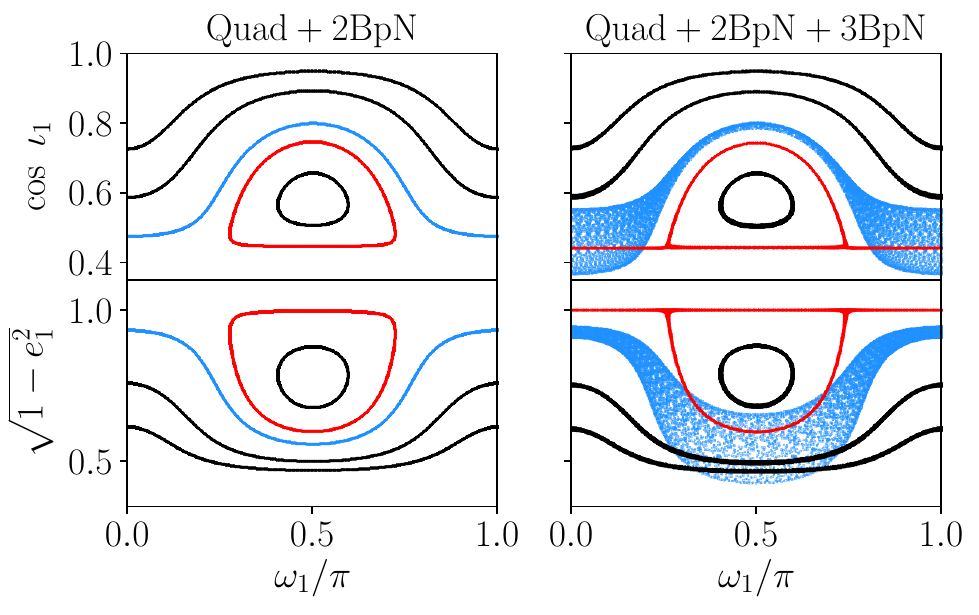}
\caption{Three-body 1pN (3BpN) effects for a circulating system, including quadrupole (Quad),inner and outer 1pN precessions (2BpN), and 3BpN effects. We plot trajectories in $(\iota_1,\omega_1)$ phase space (top) and $(e_1,\omega_1)$  phase space (bottom) for a triple with $(m_1,m_2,m_3)=(30M_\odot,20M_\odot,2\times10^7M_\odot)$, $e_2 = 0.8$, and $(a_1,\ a_2) = (0.94\ {\rm AU}, 191.86\ {\rm AU})$. Each trajectory is initialized with $\omega_1 = 90^\circ$, $\omega_2 = 120^\circ$, $\Omega_1 = 89.1^\circ$, and $\Omega_2 =  269.1^\circ$ but with different initial $e_1$ and $\iota_1$ such that $\ell_{z} = \sqrt{1-e_1} \cos\iota_1 = 0.4449$.  For circulating trajectories outside the separatrix, 3BpN effects can significantly modulate the the amplitude of the LK cycles, filling nearby regions of phase space (blue). The time evolution of the blue trajectory is plotted in Fig.~\ref{fig:timespace_circulating}. Trajectories near the separatrix switch between librating and circulating (red).}.
\label{fig:phasespace_circulating}
\end{figure}

\begin{figure}
\centering
\includegraphics[width = .45\textwidth]{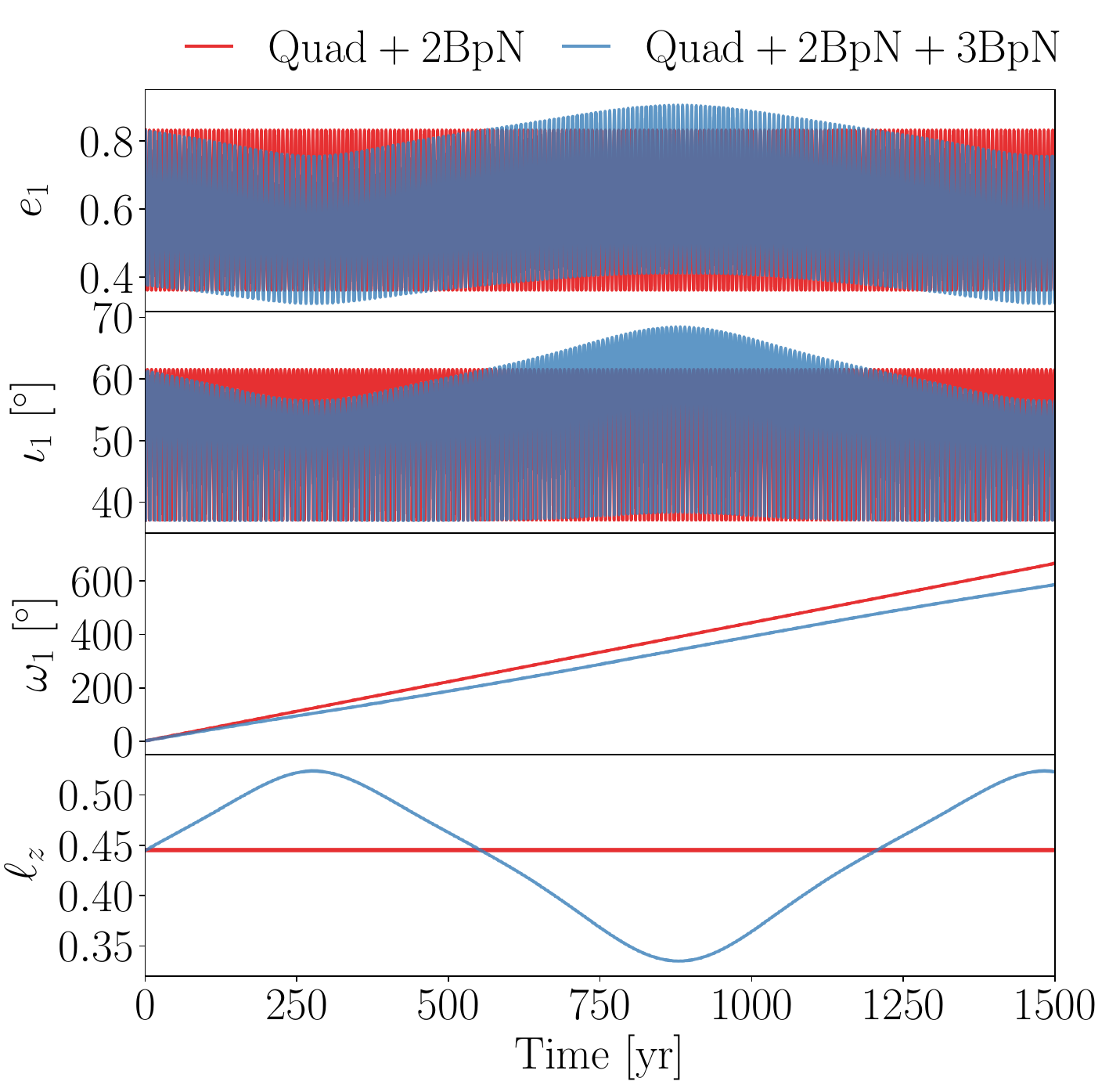}
\includegraphics[width = .48\textwidth]{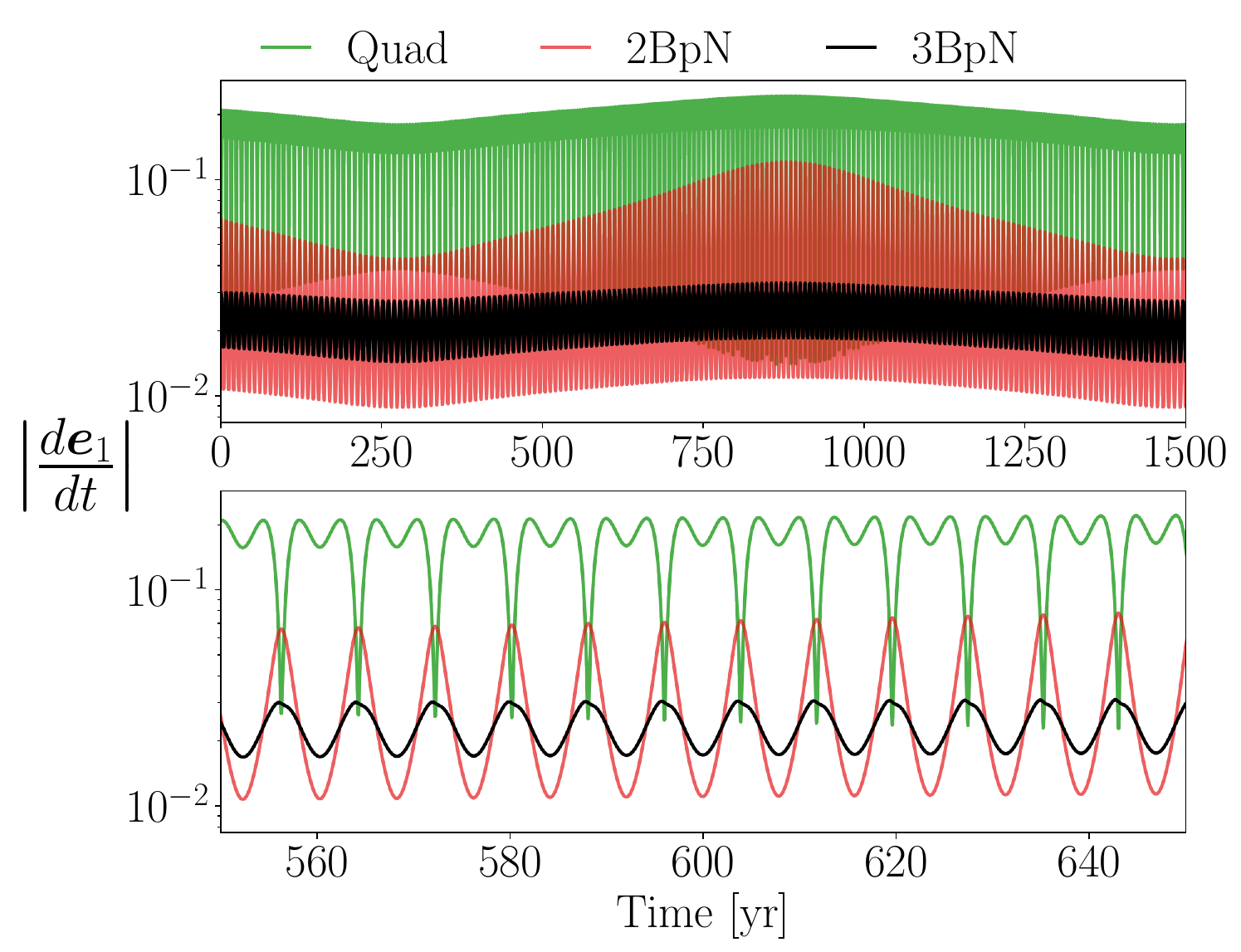}
\caption{Evolution of a triple system exhibiting the 3BpN circulating resonance.  We plot the time evolution including quadrupole and 2BpN effects (red line, top), and the time evolution including quadrupole, 2BpN, and 3BpN effects (blue line, top). We also show the instantaneous magnitudes of quadrupole, 2BpN and all 3BpN perturbations on $\dot{\bm{e}}_1$ [Eq.~(\ref{eq:edotdef})] in units of ${\rm yr}^{-1}$.} 
\label{fig:timespace_circulating}
\end{figure}

In Fig.~\ref{fig:phasespace_librating}, we plot the various parametrized trajectories in phase space each with the same initial value for
\begin{equation} \label{eq:lzdef}
    \ell_z = \sqrt{1-e_1^2}\cos\iota.
\end{equation}
When only including quadrupole and 2BpN effects, $\ell_z$ is a constant of motion in the test-particle limit ($m_2 \rightarrow 0$). Although we work outside the test-particle limit, we consider systems where the ratio of inner to outer angular momentum is sufficiently small $L_1/L_2 \sim 0.008$ so that $\ell_z$ is still nearly constant \cite{Naoz2013a}. Since $\ell_z$ is nearly constant, the trajectories are closed and exhibit either libration or circulation (see Ref.~\cite{Merritt2013DynamicsNuclei} for a review). Circulating trajectories are those for which $\omega_1$ spans all values in $(0,2\pi)$, increasing or decreasing monotonically with time. Librating trajectories are those for which $\omega_1$ spans a subset of $(0,2\pi)$, oscillating with a constant amplitude about the fixed point. The separatrix is the trajectory separating the two types of behavior.

The cross terms lead to the thickening of both librating and circulating phase space trajectories. This is due to cross-term induced oscillations in the Newtonian-order angular momentum expression, which causes $\ell_z$ to oscillate. A similar cross-term effect is described in Ref.~\cite{Will2014}. As $\ell_z$ oscillates in time, the triple's trajectory in phase space migrates through multiple nearby ``closed" trajectories corresponding to different initial $\ell_z$. For trajectories near the separatrix, this causes the system to switch between circulation and libration (red trajectory in Fig.~\ref{fig:phasespace_librating}). We note that a similar effect can be seen in triples where the octupole terms have a strong influence on the dynamics (c.f. Fig.~4 in Ref.~\cite{Ford2000}), but is identified here due to the influence of cross terms.

Maximal eccentricity growth in the inner binary occurs when 3BpN effects are comparable to 2BpN effects in magnitude. Fig.~\ref{fig:phasespace_librating} (blue trajectory) shows an example of this behavior in phase space. The addition of cross terms significantly thicken the librating trajectory, completely filling the interior region. The evolution over time for the same system is shown in Fig.~\ref{fig:timespace_librating} (top panel). The amplitude of LK oscillations in $e_1,\iota_1,\omega_1$ changes as $\ell_z$ modulates about its initial value and is larger when $\ell_z(t) < \ell_z(0)$ and smaller when $\ell_z(t) > \ell_z(0)$. We find that the opposite is true for retrograde systems.

In Fig.~\ref{fig:timespace_librating} (bottom panel), we also plot the individual contribution from each effect toward the perturbation on the inner binary's orbital vector,
\begin{equation}\label{eq:edotdef}
    \left|\frac{d \bm{e}_1}{dt}\right| = \left|\frac{d (e_1\bm{n})}{dt}\right|
\end{equation}
where $\bm{n}$ is the unit vector pointing toward the pericenter [Eq.~(\ref{eq:kepler})]. The three dominant perturbations on the inner binary are from the quadrupole (``Quad"), inner 1pN precession (``2BpN"), and de Sitter precession (``dS"). We find that significant modulations only occur, however, when including the relativistic-LK effect [Eq.~(\ref{eq:ct1p5})]. The period of LK oscillations is about $10\ {\rm yr}$, while the cross terms induce coherent modulations to the LK oscillations with a period of about $700\ {\rm yr}$. Throughout the evolution, the dS cross terms exceed the 2BpN perturbations,
\begin{equation} \label{eq:resonantcondition}
    \left|\frac{d \bm{e}_1}{dt}\right|_{\rm dS} > \left|\frac{d \bm{e}_1}{dt}\right|_{\rm 1pN}.
\end{equation}
In general, we find that in systems where Eq.~(\ref{eq:resonantcondition}) is true at some point, there is nontrivial addition of the 3BpN and 2BpN effects leading to resonantlike modulations resembling Fig.~\ref{fig:timespace_librating}. On the other hand, when the cross terms are always subdominant to 2BpN terms, the modulations are suppressed.

In Fig.~\ref{fig:phasespace_circulating}, we show an example of a second resonantlike effect for circulating trajectories. Similar to the librating behavior, the phase space trajectory is substantially thickened so the system spans a larger range of inclination and eccentricity. However, unlike the librating effects, the circulating trajectory undergoes LK oscillations where the mean eccentricity changes with $\ell_z$ and the LK oscillation amplitude is roughly constant (Fig.~\ref{fig:timespace_circulating}, top panel). During the peaks in the LK oscillations, the  dS and 1pN perturbations can exceed the quadrupole perturbations, so that
\begin{align} \label{eq:resonantcondition2}
\begin{split}
    & \left|\frac{d \bm{e}_1}{dt}\right|_{\rm 1pN}  > \left|\frac{d \bm{e}_1}{dt}\right|_{\rm quad}{\rm and}\quad \left|\frac{d \bm{e}_1}{dt}\right|_{\rm dS}  > \left|\frac{d \bm{e}_1}{dt}\right|_{\rm quad}. \\
\end{split}
\end{align}
For the system plotted in Fig.~\ref{fig:timespace_circulating} (bottom panel), this occurs during the high-eccentricity phase of the modulations, when $\ell_z < \ell_z(0)$.

\subsection{3BpN effects on a population of triples} \label{sec:population}
\begin{figure*}
\includegraphics[width = .62\textwidth]{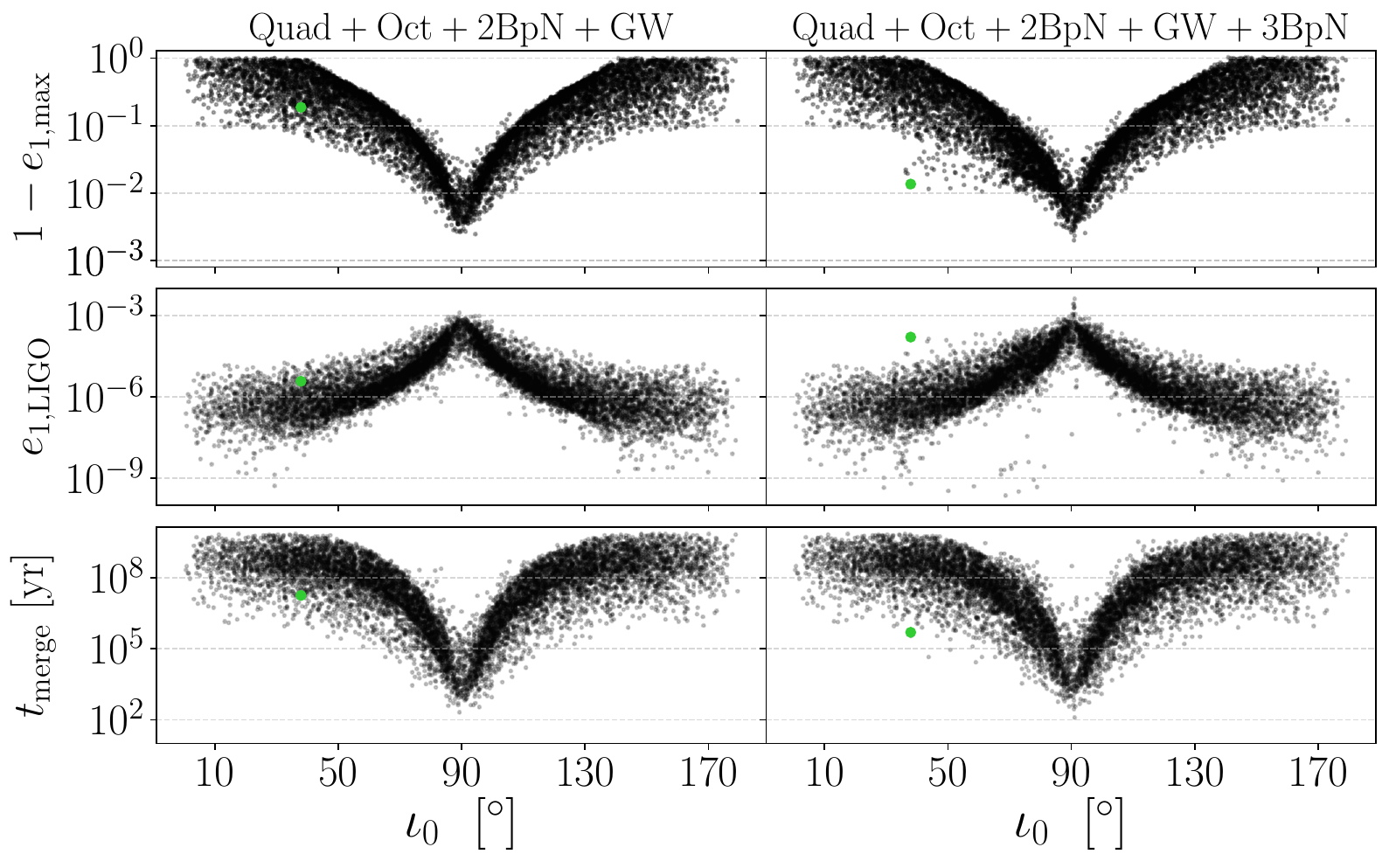}
\includegraphics[width = .36\textwidth]{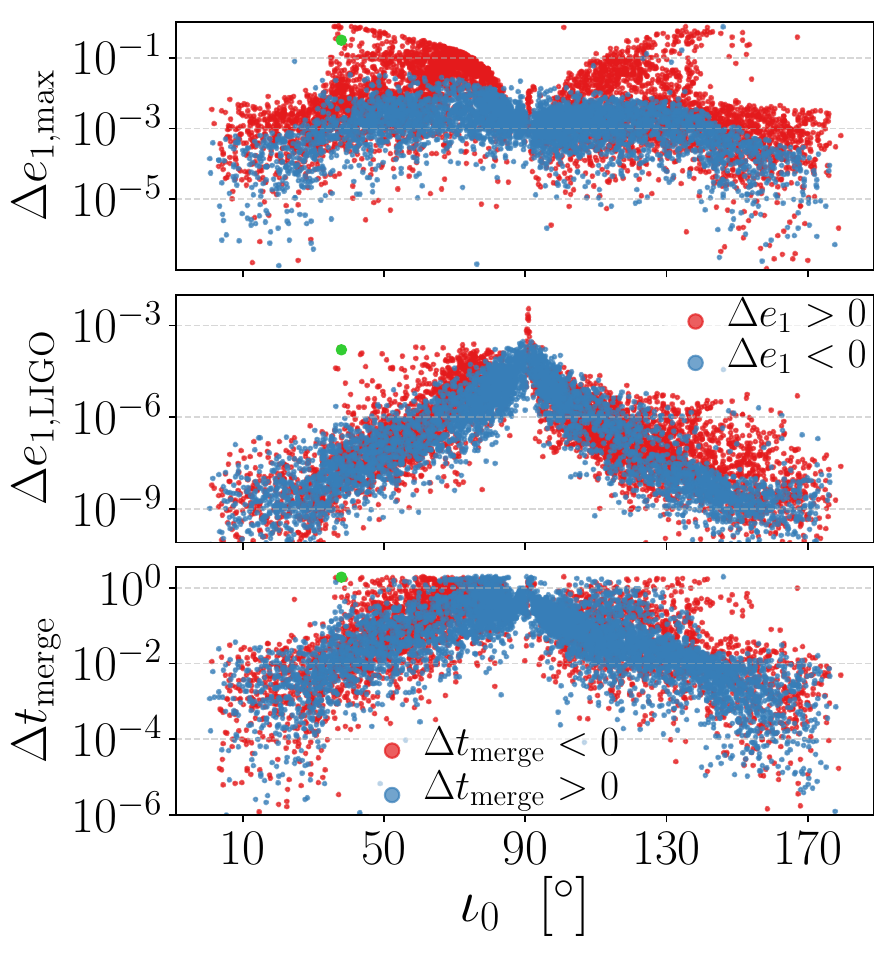}
\caption{3BpN effects on a population of merging triple systems. We plot the maximum eccentricity $e_{1,{\rm max}}$, residual eccentricity in the LIGO band $e_{1,{\rm LIGO}}$, and the merger time $t_{\rm merge}$ for evolutions with and without 3BpN effects (left). We also plot the change in each quantity after adding 3BpN effects (right). For systems with moderate inclinations $35^\circ \lesssim \iota_0 \lesssim 75^\circ$, 3BpN effects lead to enhanced eccentricities up to $e_{1,{\rm max}} \lesssim 0.99$, creating a ``shoulder"-like cluster in the final distribution. In these systems, the interaction of octupole and 3BpN effects lead to large modulations in the LK oscillations that significantly increase eccentricity. We plot the time evolution of a representative system exhibiting this mechanism in Fig.~\ref{fig:octexample}, which is also plotted above with a green dot.} 
\label{fig:butterly}
\end{figure*}

To study how 3BpN perturbations systematically affect a population of triples, we focus on the region of parameter space described in Fig.~\ref{fig:timescales}, where 3BpN effects are expected to be significant. We generate initial separations for 10,000 triples by sampling a log-uniform distribution within this $(a_1,a_2)$ region and set $e_2 = 0.8$, $m_1 = 30M_\odot$, $m_2 = 20M_\odot$ and $m_3 = 2\times10^7M_\odot$. For the inner eccentricity we assume an initially thermal distribution, uniform in $e_1^2$. We also assume an initially isotropic distribution so that $\Omega_1, \omega_j,  \cos\iota_j$ are uniformly sampled across all possible values. For each evolution we include quadrupole, octupole, and two-body 1pN secular effects on the inner and outer binary, as well as GW dissipation in the inner binary. We evolve each system twice --- with and without 3BpN cross terms. We integrate the secular equations using GSL, which implements the explicit Dormand-Prince (8,9) method with adaptive time steps \cite{Galassietal.GNUManual}. We ensure that numerical errors do not impact our overall conclusions by comparing evolutions with different error tolerances, $\epsilon_{\rm rel} = \left( 10^{-15},10^{-12}\right)$, which control the time step.

As the inner binary shrinks due to GW dissipation, it eventually enters a GW-dominated regime and decouples from the outer orbit. We integrate each system until the Keplerian orbital frequency reaches $f_{\rm orb} = 5\ {\rm Hz}$, approximately corresponding to a gravitational wave frequency of $10\ {\rm Hz}$, the lower edge of the LIGO sensitivity range, after which we consider the system ``merged". For the masses we consider, this occurs when 
\begin{equation}
    a_1 = \left(\frac{G m}{f_{\rm orb}^2}\right)^{1/3} = 4.3\times10^{-5}\ {\rm AU} = 44\ R_g,
\end{equation}
where $R_g = 2 G m/c^2$ is the gravitational radius. All systems in our population merge before a Hubble time,
\begin{equation}
    t_{\rm merge} < t_{\rm H} = 1.38\times10^{10}\ {\rm yr},
\end{equation}
which is expected given that the timescale for GW dissipation is \cite{Peters1964GravitationalMasses}
\begin{align}
\begin{split}
    t_{\rm GW} & = \frac{a_1}{ \left|\left \langle da_1/dt \right \rangle_{\rm GW}\right|} = \frac{5}{64}\frac{c^5}{G^3} \frac{a_1^4 (1-e_1)^{7/2}}{m_1 m_2 m} \\
    & = 4.3\times10^9\ {\rm yr}\times\left(\frac{a_1}{0.1{\rm AU}}\right)^4 \left(1-e_1^2\right)^{7/2}.
\end{split}
\end{align} 
For systems that achieve large eccentricities through the LK resonance, $e_1 \gtrsim 0.9$, the  merger timescale can decrease by up to three orders of magnitude. When eccentricity is very large, $e_{1,{\rm max}} \gtrsim 0.999$, the evolution becomes nonsecular, which we identify using the criterion from Ref.~\cite{Antonini2014}. We neglect these nonsecular evolutions in our analysis, which account for 4.0\% of all runs.

In Fig.~\ref{fig:butterly}, we compare the effect of 3BpN terms on the maximum eccentricity $e_{1,{\rm max}}$, the merger time $t_{\rm merge}$, and the residual eccentricity upon entering the LIGO frequency band $e_{\rm LIGO}$ as a function of initial inclination. We define $e_{\rm LIGO}$ as the eccentricity when the frequency of the peak GW harmonic reaches $10\ {\rm Hz}$:
\begin{equation}
    f_{\rm GW} = \frac{2 f_{\rm orb}(1+e_1)^{1.1954}}{(1-e_1^2)^{3/2}} = 10\ {\rm Hz}.
\end{equation}
We also plot the fractional change in merger time defined as
\begin{equation}
    \frac{\Delta t_{\rm merge}}{\left\langle t_{\rm merge} \right\rangle} = \frac{t_{\rm merge}^{\rm 3BpN} - t_{\rm merge}^{\rm 2BpN}}{\frac{1}{2}\left(t_{\rm merge}^{\rm 3BpN}+t_{\rm merge}^{\rm 2BpN}\right)}.
\end{equation}

When including 3BpN effects, a shoulderlike cluster in the maximum eccentricity distribution appears around $0.95\lesssim e_{1,{\rm max}} \lesssim 0.99$ for systems with initially moderate inclinations $35^\circ \lesssim \iota_0 \lesssim 75^\circ$, where $\iota_0$ is the initial mutual inclination. In these systems, the 3BpN effects lead to a preferential increase in $e_{1,{\rm max}}$ and $e_{1,{\rm LIGO}}$, and decrease in $t_{\rm merge}$. This effect is strongest when the 2BpN and 3BpN (de Sitter) perturbations can briefly exceed the quadrupole perturbations [Eq.~(\ref{eq:resonantcondition2})]. The resulting behavior resembles the effects shown in Fig.~\ref{fig:timespace_circulating}, where coherent perturbations to the LK oscillations occur with some characteristic amplitude and frequency. For these coherent perturbations to occur, the de Sitter term [Eq.~(\ref{eq:ct0})] must be the largest cross term, followed by the relativistic-LK terms [Eq.~(\ref{eq:ct1p5})],
\begin{equation} \label{eq:ordering}
    \left|\frac{d \bm{e}_1}{dt}\right|_{\rm RLK} <  \left|\frac{d \bm{e}_1}{dt}\right|_{\rm dS} \sim \left|\frac{d \bm{e}_1}{dt}\right|_{\rm 1pN} \lesssim \left|\frac{d \bm{e}_1}{dt}\right|_{\rm quad}.
\end{equation}

\begin{figure*}
\includegraphics[width = 1\textwidth]{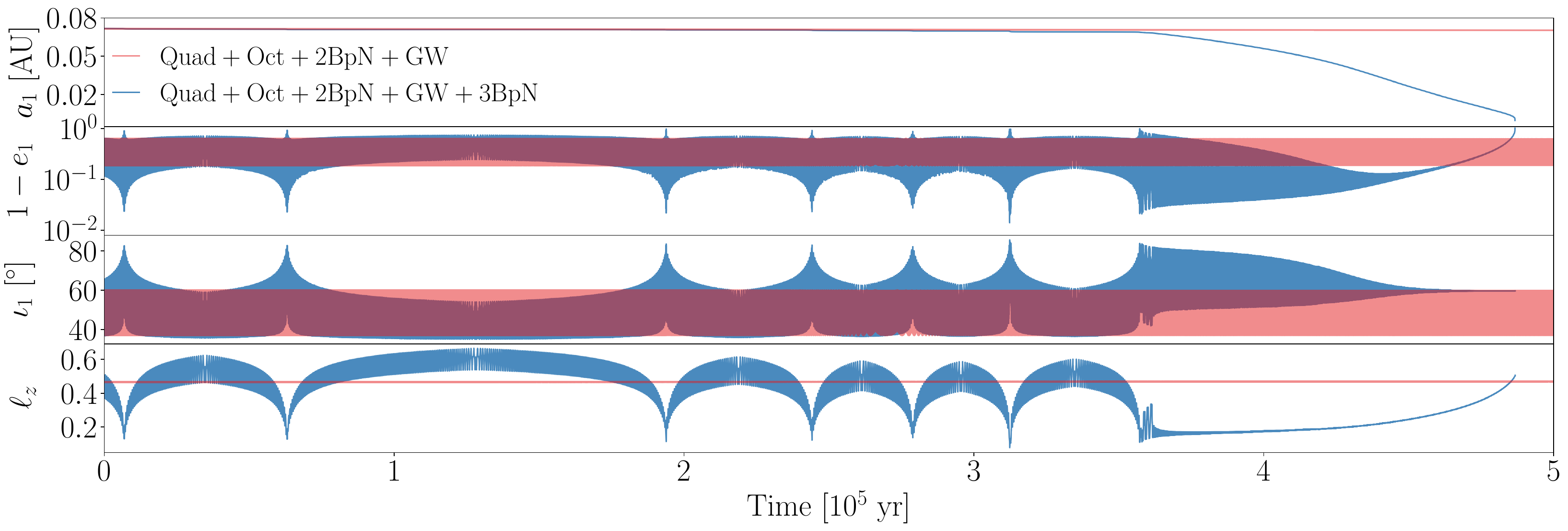}
\includegraphics[width = 1\textwidth]{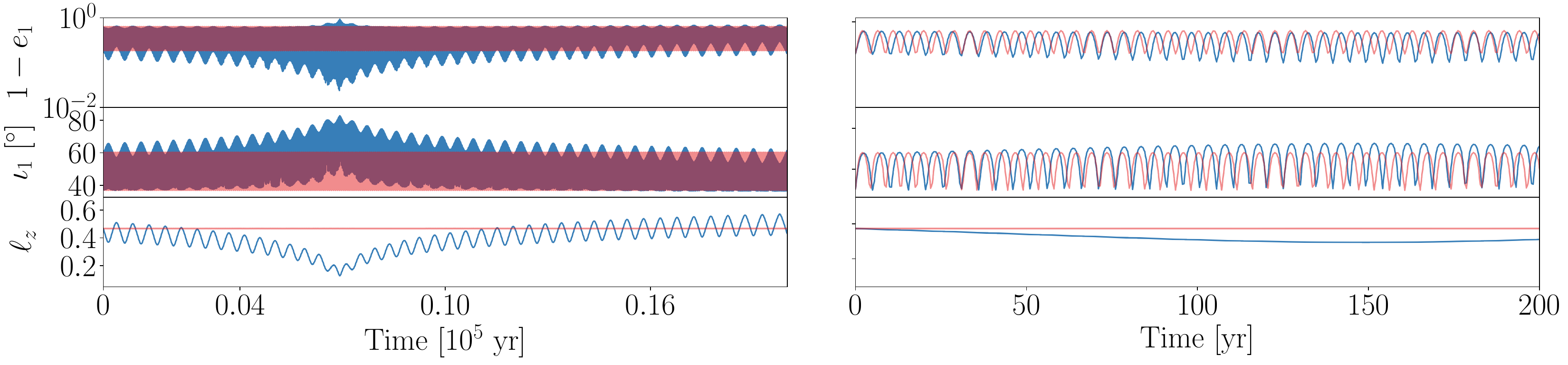}
\includegraphics[width = 1\textwidth]{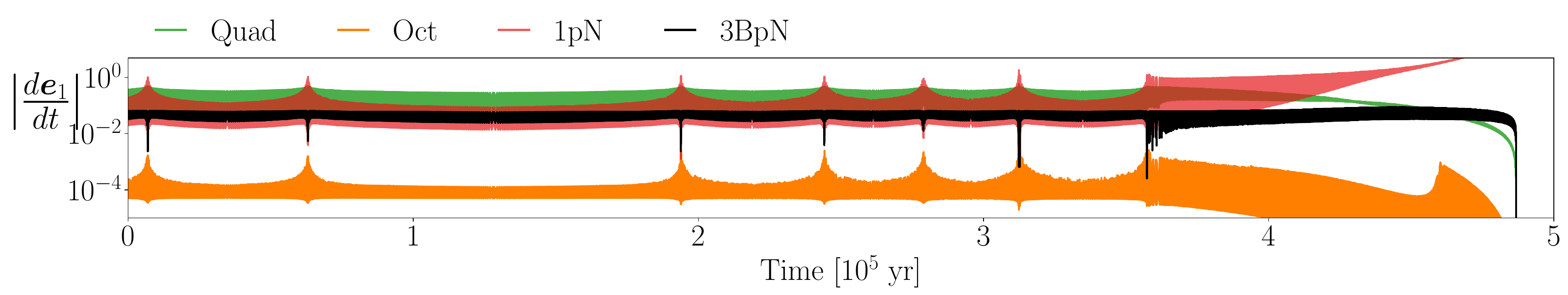}
\includegraphics[width = 1\textwidth]{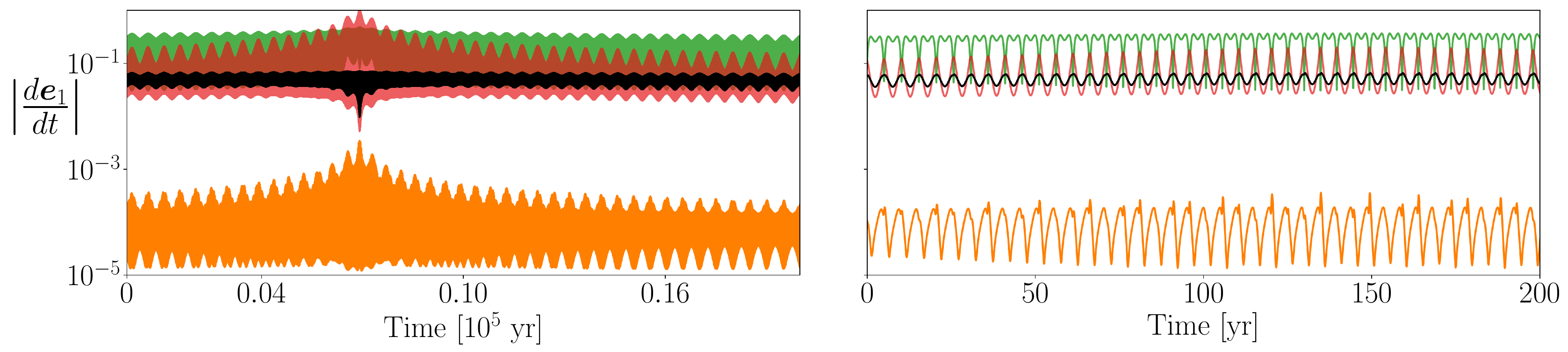}
\caption{Evolution of hierarchical triple resulting in an accelerated merger due to significant eccentricity growth from 3BpN and octupole effects. The initial parameter for this system are $(m_1,m_2,m_3) = (30M_\odot,20M_\odot,2\times 10^7 M_\odot)$, $a_1 = 0.0713\ {\rm AU}$, $e_1 = 0.808$, $a_2 = 144.694\ {\rm AU}$, and $\iota = 37.935^\circ$. For the evolution without 3BpN terms, $e_{1,{\rm max}} = 0.813$, $e_{1,{\rm LIGO}} = 3.81\times10^{-6}$ and $t_{\rm merge} = 1.76\times 10^7\ {\rm yr}$. For the evolution with 3BpN terms, $e_{1,{\rm max}} = 0.986$, $e_{1,{\rm LIGO}} = 1.63\times10^{-4}$ and $t_{\rm merge} = 4.86\times 10^5\ {\rm yr}$.}
\label{fig:octexample}
\end{figure*}

Including the octupole terms can enhance the 3BpN effects and lead to larger eccentricities than with quadrupole terms alone. We show an example of this in Fig.~\ref{fig:octexample}, also plotted with a green dot in Fig.~\ref{fig:butterly}. Initially, the system undergoes LK oscillations with a period of about $5\ {\rm yr}$. During the first few LK cycles, the eccentricity oscillates between $0.2 < e_1 < 0.8$, while the inclination oscillates between $
39^\circ < \iota_1 < 60^\circ$. The 3BpN cross terms induce periodic modulations to the LK oscillations (similar to Fig.~\ref{fig:timespace_circulating}) that occur over a period of about $300\ {\rm yr}$. Over longer timescales around $0.05-0.1\ {\rm Myr}$, the octupole terms interact with the 3BpN cross terms leading to cycles of enhanced eccentricity growth, reaching up to $e_{1,{\rm max}} = 0.986$. Eventually, the 2BpN precession arrests these octupole modulations near a phase of high eccentricity (around $t\approx 3.6\times 10^5\ {\rm yr}$) and the system transitions into a GW-dominated regime.  

For highly inclined systems $80^\circ \lesssim \iota_0 \lesssim 100^\circ$, the coherent modulations cease to be coherent if the relative ordering of the various cross terms is different from Eq.~(\ref{eq:ordering}). For instance, if the maximum eccentricity is sufficiently large $e_{1,{\rm max}} \gtrsim 0.99$, the libration cross terms, which go as $e_1 \ell_1^{-2}$ [Eq.~(\ref{eq:ctn1})], can become significant. In these systems, the cross terms lead to a systematic suppression of eccentricity growth and delayed merger times (Fig.~\ref{fig:butterly}). For moderately retrograde systems $120^\circ \lesssim \iota_0 \lesssim 160^\circ$, the 2BpN and 3BpN perturbations approach the quadrupole perturbations in magnitude, resulting in modulations with no characteristic amplitude or frequency. Although these systems may reach large eccentricities, where $e_{1,{\rm LIGO}} > 10^{-3}$ (Fig.~\ref{fig:butterly}), the long-term evolutions for these systems do not converge with our current code, unlike the coherent modulations observed in systems with initially moderate prograde inclinations (Fig.~\ref{fig:octexample}). We leave further investigation of these non-coherent behaviors to future work.

\begin{section}{Discussion} \label{sec:discussion}
In this paper we derived and investigated three-body post-Newtonian (3BpN) secular effects in hierarchical triples containing a SMBH. We expanded the Lagrange planetary equations to 1pN-octupole order with a two-parameter perturbative expansion in the pN parameter $\delta=v/c$ and the ratio of semimajor axes $\epsilon=a_1/a_2$. Using a multiple scales method, we derived secular 3BpN terms that can significantly change the evolution of the inner binary. Upon request, we will provide a \textit{Mathematica} notebook that contains a complete derivation.

When the mass of the inner binary is relatively small ($m \ll M$), three dominant 3BpN effects emerge. The main effect is the de Sitter precession, which parallel transports the inner orbit's angular momentum vector along its path around the tertiary. Other dominant effects include 1pN corrections to LK oscillations and three-body corrections to relativistic precession. While the dS term comes directly from the EIH equations, the other 3BpN terms arise from the interaction of lower-order perturbations and can be derived with a multiple-scale analysis.

For a population in the parameter space where 3BpN effects are expected to be important, we found systematic eccentricity growth for systems with initially moderate inclinations. The 3BpN effects altered the evolution of these triples by inducing coherent modulations in the quadrupole LK oscillations, which led to a larger range in eccentricity and inclination. The octupole terms enhanced the 3BpN effects inducing greater eccentricity growth, and caused systems to merge more rapidly through GW dissipation. At high inclinations, the modulations become less coherent, with varying amplitude and frequency, and can suppress eccentricity growth.

With orbit-averaged methods, one can only get an estimate on the merger times and eccentricities. It would also be insightful to compare results from this analysis with integrations from $N$-body codes that include all 3BpN effects. Given the rich and complex behavior we observe in this analysis, further work is warranted to fully explore the implications of relativistic three-body effects in hierarchical triples.

\end{section}

\section*{Acknowledgments}

Our work on this problem was supported at the Massachusetts Institute of Technology (MIT) by the National Science Foundation Grant PH-1707549. Additionally, H.\ L.\ was supported by an MIT Dean of Science Graduate Fellowship. C.\ R.\ was supported by an MIT Pappalardo Fellowship. We thank Clifford M. Will, Smadar Naoz, and Scott A. Hughes for helpful discussions related to this paper.

\bibliographystyle{apsrev4-1}

\bibliography{triples}

\appendix

\section{Appendix} \label{app:details}

We present the all the three-body pN secular terms through 1pN-quadrupole order $(\epsilon^3/c^2)$ for general masses. We also present the 1pN-octupole order $(\epsilon^4/c^2)$ terms to lowest order in $(m/M)$.

Since the equations are lengthy, we organize their presentation by distinguishing the direct cross terms which come directly from $\bm{a}_{\rm 3BpN}$, from the indirect terms which come from the interaction of lower-order corrections to $(dt/df)$, $(dt/dF)$, $P_{\rm in}$, $P_{\rm out}$, and periodic average-free perturbations $W_\alpha$.

The equations are written in terms the total inclination $\iota=\iota_1 + \iota_2$ and $\iota_1$.

\begin{widetext}

\begin{subsection}{Direct cross terms} \label{sec:DN6}
The direct cross terms come directly from $\bm{a}_{\rm 3BpN}$ [Eq.~(\ref{eq:directterms})] inserted into the planetary equations [Eq.~(\ref{eq:planetary})--(\ref{eq:planetary2})], which are then orbit averaged [Eq.~(\ref{eq:doubleorbitavg})] using the Keplerian-order expressions for $(dt/df)$, $(dt/dF)$, $P_{\rm in}$ and $P_{\rm out}$:
\begin{equation}
    \frac{d \tilde{X}_\alpha}{dt} = \frac{1}{P^{\rm K}_{\rm out}} \int\limits_0^{2\pi} \frac{1}{P_{\rm in}^{\rm K}} \int\limits_0^{2\pi} \left(\dot{X}_\alpha\right)_{\rm 3BpN} \left(\frac{dt}{df}\right)_{\rm K} \left(\frac{dt}{dF}\right)_{\rm K} df\ dF
\end{equation}

\begin{subsubsection}{1pN-quadrupole order terms}
\begin{flalign}\label{eq:dp1dtDN6}
\frac{dp_1}{dt} = \frac{33 e_1^2 G^{3/2} \sqrt{m} m_3 p_1^{3/2} \ell _2^3 \sin ^2(\iota ) \sin (2 \omega _1)}{4 c^2 p_2^3 \ell _1^2} &&
\end{flalign}

\begin{flalign}\label{eq:de1dtDN6}
\frac{de_1}{dt} = \frac{3 G^{3/2} \sqrt{m} m_3 \sqrt{p_1} (\ell _1-1){}^2 \ell _2^3 \sin ^2(\iota ) \sin (2 \omega _1) (\ell _1 ((12 \eta -23) \ell _1-22)-11)}{8 c^2 e_1^3 p_2^3} &&
\end{flalign}

\begin{flalign}\label{eq:di1dtDN6}
\frac{d\iota_1}{dt} = \frac{33 e_1^2 G^{3/2} \sqrt{m} m_3 \sqrt{p_1} \ell _2^3 \sin (2 \iota ) \sin (2 \omega _1)}{16 c^2 p_2^3 \ell _1^2} &&
\end{flalign}

\begin{flalign}\label{eq:domega1dtDN6}
\frac{d\omega_1}{dt} = &  -\frac{G^{3/2} m_3 (4m+3m_3)\ell_2^3 \csc\iota_1\sin\iota_2}{2c^2 \sqrt{M} p_2^{5/2}} &&\\\nonumber
& + \frac{G^{3/2} m_3 \ell _2^3 \csc (\iota _1) \sqrt{m p_1}}{32 c^2 e_1^4 p_2^3 \ell _1^2} \bigg( 2 (\ell _1-1){}^2 \ell _1^2 \sin (\iota _1) ((10 \eta -11) (\ell _1+1){}^2 (3 \cos (2 \iota )+1)-6 \sin ^2(\iota ) \cos (2 \omega _1) && \\\nonumber
& \times (-6 \eta +\ell _1 (6 \eta  (\ell _1-2)+5 \ell _1+34)+17))-6 e_1^4 \sin (2 \iota ) \cos (\iota _1) (-11 e_1^2 \cos (2 \omega _1)+(4 \eta -5) \ell _1^2+11) \bigg) 
\end{flalign}

\begin{flalign}\label{eq:dOmega1dtDN6}
\frac{d\Omega_1}{dt} = & -\frac{G^{3/2} m_3 (4m+3m_3)\ell_2^3 \csc\iota_1\sin\iota}{2 c^2 \sqrt{M} p_2^{5/2}} &&\\\nonumber
& + \frac{3 G^{3/2} \sqrt{m} m_3 \sqrt{p_1} \ell _2^3 \sin (2 \iota ) \csc (\iota _1) (-11 e_1^2 \cos (2 \omega _1)+(4 \eta -5) \ell _1^2+11)}{16 c^2 p_2^3 \ell _1^2} &&
\end{flalign}


\end{subsubsection}
\end{subsection}

\begin{subsubsection}{1pN-octupole order terms}
\begin{flalign}\label{eq:dp1dtDN6oct}
\frac{dp_1}{dt} = & \frac{3 e_1 e_2 \sqrt{1-4 \eta } G^{3/2} M^{3/2} p_1^2 \ell _2^3 \left(\cos (\iota ) \left(5 \cos ^2(\iota )-1\right) \sin \left(\omega _1\right) \cos \left(\omega _2\right)+\left(3-7 \cos ^2(\iota )\right) \sin \left(\omega _2\right) \cos \left(\omega _1\right)\right)}{4 c^2 p_2^{7/2} \ell _1^2} && \\\nonumber
& -\frac{15 e_1^2 G^{3/2} M^2 p_1^{5/2} \ell _2^3 \left(\sin \left(2 \omega _1\right) \left(-e_2^2 (\cos (2 \iota )+3) \cos \left(2 \omega _2\right)-12 \sin ^2(\iota )\right)+4 e_2^2 \cos (\iota ) \sin \left(2 \omega _2\right) \cos \left(2 \omega _1\right)\right)}{16 c^2 \sqrt{m} p_2^4 \ell _1^4}
\end{flalign}

\begin{flalign}\label{eq:de1dtDN6oct}
\frac{de_1}{dt} = & -\frac{3 e_2 \sqrt{1-4 \eta } G^{3/2} M^{3/2} p_1 \ell _2^3}{8 c^2 p_2^{7/2} (\ell _1+1){}^2} \bigg(\cos (\iota ) \sin (\omega _1) \cos (\omega _2) (\cos ^2(\omega _1) (5 (1-2 (\ell _1-1) \ell _1) \cos ^2(\iota )+\ell _1 (17 \ell _1+4) && \\\nonumber
& +2) + \sin ^2(\omega _1) (5 \ell _1^2 \cos ^2(\iota ) +2 (\ell _1+7) \ell _1+7)-4 (\ell _1+1){}^2)-\sin (\omega _2) \cos (\omega _1) (\cos ^2(\omega _1) ((2 \ell _1 (\ell _1+7)+7) && \\\nonumber
& \times \cos ^2(\iota )+5 \ell _1^2)+\sin ^2(\omega _1) ((\ell _1 (17 \ell _1+4)+2) \cos ^2(\iota )-10 (\ell _1-1) \ell _1+5)-4 (\ell _1+1){}^2)\bigg) && \\\nonumber
& - \frac{15 e_1 G^{3/2} M^2 p_1^{3/2} \ell _2^3 \left(\sin \left(2 \omega _1\right) \left(-e_2^2 (\cos (2 \iota )+3) \cos \left(2 \omega _2\right)-12 \sin ^2(\iota )\right)+4 e_2^2 \cos (\iota ) \sin \left(2 \omega _2\right) \cos \left(2 \omega _1\right)\right)}{32 c^2 \sqrt{m} p_2^4 \ell _1^2}
\end{flalign}

\begin{flalign}\label{eq:di1dtDN6oct}
\frac{d\iota_1}{dt} = & -\frac{3 e_1 e_2 \sqrt{1-4 \eta } G^{3/2} M^{3/2} p_1 \ell _2^3 \sin (\iota ) \left((5 \cos (2 \iota )-11) \sin \left(\omega _1\right) \cos \left(\omega _2\right)-14 \cos (\iota ) \sin \left(\omega _2\right) \cos \left(\omega _1\right)\right)}{16 c^2 p_2^{7/2} \ell _1^2} && \\\nonumber
& -\frac{3 G^{3/2} M^2 p_1^{3/2} \ell _2^3 \sin (\iota ) \left(e_2^2 \sin \left(2 \omega _2\right) \left(-5 e_1^2 \cos \left(2 \omega _1\right)+3 \ell _1^2-5\right)-5 e_1^2 \cos (\iota ) \sin \left(2 \omega _1\right) \left(6-e_2^2 \cos \left(2 \omega _2\right)\right)\right)}{16 c^2 \sqrt{m} p_2^4 \ell _1^4}
\end{flalign}

\begin{flalign}\label{eq:domega1dtDN6oct}
\frac{d\omega_1}{dt} = & -\frac{3 e_2 \sqrt{1-4 \eta } G^{3/2} M^{3/2} p_1 \ell _2^3}{8 c^2 p_2^{7/2} \sqrt{1-\ell _1} \ell _1^2 (\ell _1+1){}^{5/2}} \bigg (\cos (\iota ) (\frac{1}{2} \cos (\omega _1) \cos (\omega _2) (-5 (e_1^2+2 \ell _1^4-4 \ell _1^3+2 \ell _1) \cos (2 \omega _1)+20 \ell _1^4 && \\\nonumber
& +20 \ell _1^3+\ell _1^2-18 \ell _1-9)-17 (\ell _1-1) (\ell _1+1){}^3 \sin (\iota ) \cot (\iota _1) \sin (\omega _1) \sin (\omega _2))+\cos ^2(\iota ) \sin (\omega _1) \sin (\omega _2) ((\ell _1 (\ell _1 && \\\nonumber
& \times (\ell _1 (17 \ell _1+44)+7)-30)-15) \sin ^2(\omega _1)+(\ell _1 (3 \ell _1-2) (\ell _1 (9 \ell _1+14)+10)-10) \cos ^2(\omega _1))+ 5 \cos ^3(\iota ) && \\\nonumber
& \times \cos (\omega _1) \cos (\omega _2) ((\ell _1 (-\ell _1^3+4 \ell _1^2+\ell _1-2)-1) \sin ^2(\omega _1)+\ell _1^4 \cos ^2(\omega _1))-7 (\ell _1-1) (\ell _1+1){}^3 \sin (\iota ) \cot (\iota _1) && \\\nonumber
& \times \cos (\omega _1) \cos (\omega _2)+\sin (\omega _1) \sin (\omega _2) (5 \ell _1^4 \sin ^2(\omega _1)-5 (\ell _1 (\ell _1 ((\ell _1-4) \ell _1-1)+2)+1) \cos ^2(\omega _1)-4 (\ell _1+1){}^2 (3 \ell _1^2-2))\bigg) && \\\nonumber
& -\frac{3 G^{3/2} M^2 p_1^{3/2} \ell _2^3 }{32 c^2 \sqrt{m} p_2^4 \ell _1^4}\bigg(10 e_1^2 e_2^2 \sin (\iota ) \cot (\iota _1) \sin (2 \omega _1) \sin (2 \omega _2)-\ell _1^2 \cos (2 \iota ) (5 \cos (2 \omega _1)-3) (6-e_2^2 \cos (2 \omega _2)) && \\\nonumber
& +20 e_2^2 \ell _1^2 \cos (\iota ) \sin (2 \omega _1) \sin (2 \omega _2)+\sin (2 \iota ) \cot (\iota _1) (6-e_2^2 \cos (2 \omega _2)) (-5 e_1^2 \cos (2 \omega _1)-3 \ell _1^2+5)-3 \ell _1^2 (5 \cos (2 \omega _1) && \\\nonumber
& +1) (e_2^2 (-\cos (2 \omega _2))-2)\bigg)
\end{flalign}

\begin{flalign}\label{eq:dOmega1dtDN6oct}
\frac{d\Omega_1}{dt} = & \frac{3 e_1 e_2 \sqrt{1-4 \eta } G^{3/2} M^{3/2} p_1 \ell _2^3 \sin (\iota ) \csc \left(\iota _1\right) \left(17 \cos (\iota ) \sin \left(\omega _1\right) \sin \left(\omega _2\right)+7 \cos \left(\omega _1\right) \cos \left(\omega _2\right)\right)}{8 c^2 p_2^{7/2} \ell _1^2} && \\\nonumber
& + \frac{3 G^{3/2} M^2 p_1^{3/2} \ell _2^3 \sin (\iota ) \csc \left(\iota _1\right) \left(5 e_1^2 e_2^2 \sin \left(2 \omega _1\right) \sin \left(2 \omega _2\right)+\cos (\iota ) \left(6-e_2^2 \cos \left(2 \omega _2\right)\right) \left(-5 e_1^2 \cos \left(2 \omega _1\right)-3 \ell _1^2+5\right)\right)}{16 c^2 \sqrt{m} p_2^4 \ell _1^4}
\end{flalign}

\end{subsubsection}

\begin{subsection}{Indirect cross terms due to corrections to $(dt/df)$, $(dt/dF)$, $P_{\rm in}$, and $P_{\rm out}$} \label{sec:ID6}
These cross terms come from 1pN and quadrupole corrections to $(dt/df)$, $(dt/dF)$, $P_{\rm in}$, and $P_{\rm out}$, which combine with perturbations from $\bm{a}_{\rm quad}$ and $\bm{a}_{\rm 1pN}$. 

\begin{subsubsection}{Cross terms from $(dt/dF)_{\rm 1pN} \times (\dot{X}_\alpha )_{\rm quad}$}
These cross terms come from 1pN corrections to $(dt/dF)$ which combine with $(\dot{X}_\alpha )_{\rm quad}$:
\begin{equation}
    \frac{d \tilde{X}_\alpha}{dt} = \frac{1}{P^{\rm K}_{\rm out}} \int\limits_0^{2\pi} \frac{1}{P_{\rm in}^{\rm K}} \int\limits_0^{2\pi} \left(\dot{X}_\alpha\right)_{\rm quad} \left(\frac{dt}{df}\right)_{\rm K} \left(\frac{dt}{dF}\right)_{\rm 1pN} df\ dF.
\end{equation}
\begin{flalign}\label{eq:dp1dtID6a}
\frac{dp_1}{dt} = & \frac{15 e_1^2 G^{3/2} M^2 p_1^{5/2} \ell _2^3}{32 c^2 \sqrt{m} p_2^4 \ell _1^4}\bigg(\sin (2 \omega _1) ((\ell _2^2+12) (\cos (2 \iota )+3) \cos (2 \omega _2)+4 (\ell _2^2-4) \sin ^2(\iota ))-4 (\ell _2^2+12) \cos (\iota ) && \\\nonumber
& \times \sin (2 \omega _2) \cos (2 \omega _1)\bigg)
\end{flalign}

\begin{flalign}\label{eq:de1dtID6a}
\frac{de_1}{dt} = & \frac{15 e_1 M^2 \ell _2^3 (G p_1){}^{3/2} }{64 c^2 \sqrt{m} p_2^4 \ell _1^2} \bigg (\sin (2 \omega _1) ((\ell _2^2+12) (\cos (2 \iota )+3) \cos (2 \omega _2)+4 (\ell _2^2-4) \sin ^2(\iota ))-4 (\ell _2^2+12) && \\\nonumber
& \times \cos (\iota ) \sin (2 \omega _2) \cos (2 \omega _1)\bigg)
\end{flalign}

\begin{flalign}\label{eq:di1dtID6a}
\frac{d\iota_1}{dt} = & -\frac{3 M^2 \ell _2^3 (G p_1){}^{3/2} \sin (\iota )}{32 c^2 \sqrt{m} p_2^4 \ell _1^4} \bigg(5 e_1^2 \cos (\iota ) \sin (2 \omega _1) ((\ell _2^2+12) \cos (2 \omega _2)-2 \ell _2^2+8)+(\ell _2^2+12) \sin (2 \omega _2) (5 (\ell _1^2 && \\\nonumber
& -1) \cos (2 \omega _1)+3 \ell _1^2-5)\bigg)
\end{flalign}

\begin{flalign}\label{eq:domega1dtID6a}
\frac{d\omega_1}{dt} = & -\frac{3 M^2 \ell _2^3 (G p_1){}^{3/2} }{16 c^2 \sqrt{m} p_2^4 \ell _1^4} \bigg(\sin ^2(\omega _1) (\cos ^2(\omega _2) ((\ell _2^2-20) \cos (\iota ) (4 \ell _1^2 \cos (\iota )+(5-4 \ell _1^2) \sin (\iota ) \cot (\iota _1))-\ell _1^2 (3 \ell _2^2 && \\\nonumber
& +4))+\sin ^2(\omega _2) ((3 \ell _2^2+4) \cos (\iota ) (4 \ell _1^2 \cos (\iota )+(5-4 \ell _1^2) \sin (\iota ) \cot (\iota _1))-\ell _1^2(\ell _2^2-20)))+\ell _1^2 \cos ^2(\omega _1) (\cos ^2(\omega _2) && \\\nonumber
& \times ((\ell _2^2-20) \cos (\iota ) (\sin (\iota ) \cot (\iota _1)-\cos (\iota ))+4 (3 \ell _2^2+4))+\sin ^2(\omega _2) ((3 \ell _2^2+4) \cos (\iota ) (\sin (\iota ) \cot (\iota _1)-\cos (\iota )) && \\\nonumber
& +4 (\ell _2^2-20)))+10 (\ell _2^2+12) \sin (\omega _2) \sin (\omega _1) \cos (\omega _1) \cos (\omega _2) (2 \ell _1^2 \cos (\iota )-(\ell _1^2-1) \sin (\iota ) \cot (\iota _1))-4 \ell _1^2 (\ell _2^2-4)\bigg)
\end{flalign}

\begin{flalign}\label{eq:dOmega1dtID6a}
\frac{d\Omega_1}{dt} = & \frac{3 M^2 \ell _2^3 (G p_1){}^{3/2} \sin (\iota ) \csc (\iota _1)}{16 c^2 \sqrt{m} p_2^4 \ell _1^4} \bigg(5 e_1^2 (\ell _2^2+12) \sin (\omega _1) \sin (2 \omega _2) \cos (\omega _1)-\cos (\iota ) ((\ell _2^2+12) \cos (2 \omega _2)-2 \ell _2^2 && \\\nonumber
& +8) ((5-4 \ell _1^2) \sin ^2(\omega _1)+\ell _1^2 \cos ^2(\omega _1))\bigg)
\end{flalign}

\end{subsubsection}

\begin{subsubsection}{Cross terms from $(dt/dF)_{\rm quad} \times (\dot{X}_\alpha )_{\rm 1pN}$}
These cross terms come from quadrupole corrections to $(dt/dF)$ which combine with $(\dot{X}_\alpha )_{\rm 1pN}$:
\begin{equation}
    \frac{d \tilde{X}_\alpha}{dt} = \frac{1}{P^{\rm K}_{\rm out}} \int\limits_0^{2\pi} \frac{1}{P_{\rm in}^{\rm K}} \int\limits_0^{2\pi} \left(\dot{X}_\alpha\right)_{\rm 1pN} \left(\frac{dt}{df}\right)_{\rm K} \left(\frac{dt}{dF}\right)_{\rm quad} df\ dF.
\end{equation}
\begin{flalign}\label{eq:dp1dtID6b}
\frac{dp_1}{dt} = & \frac{6 (\eta -2) \eta  G m (\ell _1-1){}^2 \ell _1^2 (\ell _2-1){}^2 \ell _2^3 \sqrt{G m p_1}}{c^2 e_1^2 e_2^4 p_2^2} \bigg((\cos (2 \iota )+3) \sin (\omega _1) \cos (\omega _1) \cos (2 \omega _2)-2 \cos (\iota ) \sin (2 \omega _2) && \\\nonumber
& \times \cos (2 \omega _1)\bigg)
\end{flalign}

\begin{flalign}\label{eq:de1dtID6b}
\frac{de_1}{dt} = & -\frac{3 \eta  (\ell _1-1){}^2 \ell _1^3 (\ell _2-1){}^2 \ell _2^3 (G m)^{3/2} (-5 \eta +11 \eta  \ell _1-14 \ell _1+10)}{4 c^2 e_1^3 e_2^4 \sqrt{p_1} p_2^2}\bigg((\cos (2 \iota )+3) \sin (\omega _1) \cos (\omega _1) \cos (2 \omega _2) && \\\nonumber
& -2 \cos (\iota ) \sin (2 \omega _2) \cos (2 \omega _1)\bigg)
\end{flalign}

\begin{flalign}\label{eq:domega1dtID6b}
\frac{d\omega_1}{dt} = & \frac{3 \eta  (\ell _1-1){}^2 \ell _1^2 (\ell _2-1){}^2 \ell _2^3 (G m)^{3/2}}{64 c^2 e_1^4 e_2^4 \sqrt{p_1} p_2^2} \bigg(4 (\cos (2 \iota )+3) \cos (2 \omega _1) \cos (2 \omega _2) (7 \eta +\ell _1 (-22 \eta +(3 \eta +2) \ell _1+28) && \\\nonumber
& -22)-8 (\ell _1+1) \sin ^2(\iota ) \cos (2 \omega _2) (-7 \eta +3 \eta  \ell _1+2 \ell _1+22)+16 \cos (\iota ) \sin (2 \omega _1) \sin (2 \omega _2) (7 \eta +\ell _1 (-22 \eta && \\\nonumber
& + 3 \eta  \ell _1+2 \ell _1+28)-22)\bigg)
\end{flalign}

\end{subsubsection}

\begin{subsubsection}{Cross terms from $(dt/df)_{\rm 1pN} \times (\dot{X}_\alpha )_{\rm quad}$}
These cross terms come from 1pN corrections to $(dt/df)$ which combine with $(\dot{X}_\alpha )_{\rm quad}$:
\begin{equation}
    \frac{d \tilde{X}_\alpha}{dt} = \frac{1}{P^{\rm K}_{\rm out}} \int\limits_0^{2\pi} \frac{1}{P_{\rm in}^{\rm K}} \int\limits_0^{2\pi} \left(\dot{X}_\alpha\right)_{\rm quad} \left(\frac{dt}{df}\right)_{\rm 1pN} \left(\frac{dt}{dF}\right)_{\rm K} df\ dF
\end{equation}
\begin{flalign}\label{eq:dp1dtID6c}
\frac{dp_1}{dt} = & \frac{3 G^{3/2} \sqrt{m} m_3 p_1^{3/2} \ell _2^3 \sin ^2(\iota ) \sin (2 \omega _1) (\eta -3 (\eta +2) \ell _1^6+16 (\eta -5) \ell _1^5-17 (\eta -10) \ell _1^4+(3 \eta -122) \ell _1^2+38)}{8 c^2 e_1^4 p_2^3 \ell _1^2}
\end{flalign}

\begin{flalign}\label{eq:de1dtID6c}
\frac{de_1}{dt} = & \frac{3 G^{3/2} \sqrt{m} m_3 \sqrt{p_1} (\ell _1-1){}^2 \ell _2^3 \sin ^2(\iota ) \sin (2 \omega _1) }{16 c^2 e_1^5 p_2^3} \bigg(\eta  (\ell _1 (3 \ell _1 (\ell _1 (3 \ell _1-8)+10)-2)-1)+2 \ell _1 (\ell _1 (\ell _1 && \\\nonumber
& \times (5 \ell _1+66)-32)-38)-38\bigg)
\end{flalign}

\begin{flalign}\label{eq:di1dtID6c}
\frac{d\iota_1}{dt} = & \frac{3 G^{3/2} \sqrt{m} m_3 \sqrt{p_1} \ell _2^3 \sin (2 \iota ) \sin (2 \omega _1) (\eta -3 (\eta +2) \ell _1^6+16 (\eta -5) \ell _1^5-17 (\eta -10) \ell _1^4+(3 \eta -122) \ell _1^2+38)}{32 c^2 e_1^4 p_2^3 \ell _1^2}
\end{flalign}

\begin{flalign}\label{eq:domega1dtID6c}
\frac{d\omega_1}{dt} = & \frac{G^{3/2} \sqrt{m} m_3 \sqrt{p_1} (\ell _1-1){}^2 \ell _2^3}{32 c^2 e_1^6 p_2^3 \ell _1^2}  \bigg(-6 e_1^2 \sin (\iota ) \cos (\iota ) \cot (\iota _1) \cos (2 \omega _1) (-\eta +\ell _1 (\ell _1 (-6 \eta +\ell _1 (-10 \eta +3 (\eta +2) && \\\nonumber
& \times \ell _1 +92)+8)-2 (\eta +38))-38)+6 \ell _1^2 \sin ^2(\iota ) \cos (2 \omega _1) (\eta  ((\ell _1-2) \ell _1 (2 \ell _1 (3 \ell _1+8)-17)+17)+2 \ell _1 (\ell _1 (2 \ell _1 && \\\nonumber
& \times (\ell _1+14)+63)-74)-74)+\ell _1^2 (\ell _1+1) (3 \cos (2 \iota )+1) (3 (\eta -10)+\ell _1 (3 (\eta -10)+2 \ell _1 (-4 \eta +(3 \eta +2) \ell _1 && \\\nonumber
& +22)))-3 (\ell _1-1) (\ell _1+1){}^3 \sin (2 \iota ) \cot (\iota _1) (-\eta +(9 \eta +2) \ell _1^2-38)\bigg)
\end{flalign}

\begin{flalign}\label{eq:dOmega1dtID6c}
\frac{d\Omega_1}{dt} = & \frac{3 G^{3/2} \sqrt{m} m_3 \sqrt{p_1} \ell _2^3 \sin (\iota ) \cos (\iota ) \csc (\iota _1)}{16 c^2 e_1^4 p_2^3 \ell _1^2}\bigg((\ell _1-1){}^2 \cos (2 \omega _1) (-\eta +\ell _1 (\ell _1 (-6 \eta +\ell _1 (-10 \eta +3 (\eta +2) \ell _1 && \\\nonumber
& +92)+8)-2 (\eta +38))-38)-e_1^4 (-\eta +(9 \eta +2) \ell _1^2-38)\bigg)
\end{flalign}

\end{subsubsection}

\begin{subsubsection}{Cross terms from $(dt/df)_{\rm quad} \times (\dot{X}_\alpha )_{\rm 1pN}$}
These cross terms come from quadrupole corrections to $(dt/df)$ which combine with $(\dot{X}_\alpha )_{\rm 1pN}$:
\begin{equation}
    \frac{d \tilde{X}_\alpha}{dt} = \frac{1}{P^{\rm K}_{\rm out}} \int\limits_0^{2\pi} \frac{1}{P_{\rm in}^{\rm K}} \int\limits_0^{2\pi} \left(\dot{X}_\alpha\right)_{\rm 1pN} \left(\frac{dt}{df}\right)_{\rm quad} \left(\frac{dt}{dF}\right)_{\rm K} df\ dF
\end{equation}
\begin{flalign}\label{eq:dp1dtID6d}
\frac{dp_1}{dt} = & \frac{3 (\eta -2) G^{3/2} \sqrt{m} m_3 p_1^{3/2} ((\ell _1 (\ell _1+4)-9) \ell _1^4+5 \ell _1^2-1) \ell _2^3 \sin ^2(\iota ) \sin (2 \omega _1)}{2 c^2 e_1^4 p_2^3 \ell _1^2}
\end{flalign}

\begin{flalign}\label{eq:de1dtID6d}
\frac{de_1}{dt} = & -\frac{3 G^{3/2} \sqrt{m} m_3 \sqrt{p_1} (\ell _1-1){}^2 \ell _2^3 \sin ^2(\iota ) \sin (2 \omega _1)}{16 c^2 e_1^5 p_2^3} \bigg(-\eta +\eta  \ell _1 (\ell _1 (\ell _1 (5 \ell _1+28)-42)-2)-2 \ell _1 (\ell _1 (\ell _1 && \\\nonumber
& \times (7 \ell _1+22)-56)+6)-6\bigg)
\end{flalign}

\begin{flalign}\label{eq:domega1dtID6d}
\frac{d\omega_1}{dt} = & \frac{G^{3/2} \sqrt{m} m_3 \sqrt{p_1} (\ell _1-1){}^2 \ell _2^3}{32 c^2 e_1^6 p_2^3} \bigg(6 \sin ^2(\iota ) \cos (2 \omega _1) (\eta  ((\ell _1-2) \ell _1 (2 \ell _1 (3 \ell _1+8)-17)+17)+2 \ell _1 (\ell _1 (2 \ell _1 && \\\nonumber
& \times (\ell _1+14)+63)-74)-74)+(\ell _1+1) (3 \cos (2 \iota )+1) (3 (\eta -10)+\ell _1 (3 (\eta -10)+2 \ell _1 (-4 \eta +(3 \eta +2) \ell _1+22)))\bigg)
\end{flalign}
\end{subsubsection}

\begin{subsubsection}{Cross terms from $P_{\rm in}^{\rm 1pN} \times (\dot{X}_\alpha )_{\rm quad}$}
These cross terms come from 1pN corrections to $P_{\rm in}$ which combine with $(\dot{X}_\alpha )_{\rm quad}$:
\begin{equation} \label{eq:crossPin1pN}
    \frac{d \tilde{X}_\alpha}{dt} = \frac{1}{P^{\rm K}_{\rm out}} \int\limits_0^{2\pi} -\frac{P_{\rm in}^{\rm 1pN}}{(P_{\rm in}^{\rm K})^2} \int\limits_0^{2\pi} \left(\dot{X}_\alpha\right)_{\rm quad} \left(\frac{dt}{df}\right)_{\rm K} \left(\frac{dt}{dF}\right)_{\rm K} df\ dF,
\end{equation}
where above we expanded $1/P_{\rm in}$ [Eq.~(\ref{eq:singleorbitavg})] to linear order in $P_{\rm in}^{\rm 1pN}$ and
\begin{equation}
    P_{\rm in}^{\rm 1pN} = \int\limits_0^{2\pi} \left(\frac{dt}{df}\right)_{\rm 1pN} df = P_{\rm in}^{\rm K} \left(\frac{G m}{p_1 c^2}\right) \frac{e_1^2 (-(21 \eta +8)) \left(\ell _1+1\right)+21 \eta +\ell _1 \left(21 \eta +\ell _1 \left(80-\eta  \left(9 \ell _1+49\right)\right)+8\right)+8}{1+\ell_1}.
\end{equation}
There are no contributions to $P^{\rm 1pN}_{\rm in}$ due to periodic perturbations since $P_{\rm in}^{\rm K}$ only depends on the elements $e_1$ and $p_1$, which are not perturbed at 1pN order. $P_{\rm in}^{\rm 1pN}$ does not depend on $F$ and can be factored outside the outer orbit integral [Eq.~(\ref{eq:crossPin1pN})]. As a result, these cross terms are equal to the usual secular quadrupole terms times a multiplicative factor:
\begin{equation}
    \frac{d \tilde{X}_\alpha}{dt} = -\frac{P_{\rm in}^{\rm 1pN}}{P_{\rm in}^{\rm K}} \left(\frac{d \tilde{X}_\alpha}{dt}\right)_{\rm quad}
\end{equation}
where the secular quadrupole terms can be found in the literature (e.g. Refs.~\cite{Will2017a,Liu2015}).
\end{subsubsection}

\begin{subsubsection}{Cross terms from $P_{\rm in}^{\rm quad} \times (\dot{X}_\alpha )_{\rm 1pN}$}
These cross terms come from quadrupole corrections to $P_{\rm in}$ which combine with $(\dot{X}_\alpha )_{\rm 1pN}$:
\begin{equation}
    \frac{d \tilde{X}_\alpha}{dt} = \frac{1}{P^{\rm K}_{\rm out}} \int\limits_0^{2\pi} -\frac{P_{\rm in}^{\rm quad}}{(P_{\rm in}^{\rm K})^2} \int\limits_0^{2\pi} \left(\dot{X}_\alpha\right)_{\rm 1pN} \left(\frac{dt}{df}\right)_{\rm K} \left(\frac{dt}{dF}\right)_{\rm K} df\ dF,
\end{equation}
where above we expanded $1/P_{\rm in}$ [Eq.~(\ref{eq:singleorbitavg})] to linear order in $P_{\rm in}^{\rm quad}$ and
\begin{flalign}
    P_{\rm in}^{\rm quad} = & \int\limits_0^{2\pi} \left \lbrack \left(\frac{dt}{df}\right)_{\rm quad} + W_\beta^{30}\frac{\partial}{\partial \tilde{X}_\beta} \left(\frac{dt}{df}\right)_{\rm K} \right \rbrack df && \\\nonumber
    = & P_{\rm in}^{\rm K} \frac{m_3 (1+e_2 \cos(F))^3 p_1^3}{64 m p_2^3 \ell_1^6} \bigg(-5 (3 \ell _1^2-7) (6 \cos (2 \iota ) \sin ^2(F+\omega _2)+3 \cos (2 (F+\omega _2))+1)-3 (17 \ell _1^2-49) && \\\nonumber
    & \times \cos (2 \omega _1) (-2 \cos (2 \iota ) \sin ^2(F+\omega _2)+3 \cos (2 (F+\omega _2))+1)+12 (49-17 \ell _1^2) \cos (\iota ) \sin (2 \omega _1) \sin (2 (F+\omega _2)) \bigg).
\end{flalign}
The periodic contributions average to zero, so the only correction comes from $(dt/df)_{\rm quad}$, leading to
\begin{flalign}\label{eq:domega1dtID6f}
\frac{d\omega_1}{dt} = & \frac{3 G^{3/2} m_3 \ell _2^3 \sqrt{m p_1} \left(6 \left(17 \ell _1^2-49\right) \sin ^2(\iota ) \cos \left(2 \omega _1\right)+5 \left(3 \ell _1^2-7\right) (3 \cos (2 \iota )+1)\right)}{64 c^2 p_2^3 \ell _1^3}.
\end{flalign}
\end{subsubsection}

\begin{subsubsection}{Cross terms from $P_{\rm out}^{\rm 1pN} \times (\dot{X}_\alpha )_{\rm quad}$}
These cross terms come from 1pN corrections to $P_{\rm out}$ which combine with $(\dot{X}_\alpha )_{\rm quad}$:
\begin{equation} \label{eq:crossPinquad}
    \frac{d \tilde{X}_\alpha}{dt} = -\frac{P_{\rm out}^{\rm 1pN}}{(P_{\rm out}^{\rm K})^2} \int\limits_0^{2\pi} \frac{1}{P_{\rm in}^{\rm K}} \int\limits_0^{2\pi} \left(\dot{X}_\alpha\right)_{\rm quad} \left(\frac{dt}{df}\right)_{\rm K} \left(\frac{dt}{dF}\right)_{\rm K} df\ dF,
\end{equation}
where
\begin{flalign}
    P_{\rm out}^{\rm 1pN} & = \int\limits_0^{2\pi} \left \lbrack \left(\frac{dt}{dF}\right)_{\rm 1pN} + W_\beta^{\frac{5}{2}1}\frac{\partial}{\partial \tilde{X}_\beta} \left(\frac{dt}{dF}\right)_{\rm K} \right \rbrack dF = P_{\rm out}^{\rm K} \left( \frac{3 GM}{2 c^2 p_2} \right) \frac{(-3+\ell_2^2)(-5+2\ell_2^2)}{\ell_2^2}.
\end{flalign}
These cross terms are equal to the usual secular quadrupole terms times a multiplicative factor:
\begin{equation}
    \frac{d \tilde{X}_\alpha}{dt} = -\frac{P_{\rm out}^{\rm 1pN}}{P_{\rm in}^{\rm K}} \left(\frac{d \tilde{X}_\alpha}{dt}\right)_{\rm quad}
\end{equation}
\end{subsubsection}

\begin{subsubsection}{Cross terms from $P_{\rm out}^{\rm quad} \times (\dot{X}_\alpha )_{\rm 1pN}$}
These cross terms come from quadrupole corrections to $P_{\rm out}$ which combine with $(\dot{X}_\alpha )_{\rm 1pN}$:
\begin{equation} 
    \frac{d \tilde{X}_\alpha}{dt} = -\frac{P_{\rm out}^{\rm quad}}{(P_{\rm out}^{\rm K})^2} \int\limits_0^{2\pi} \frac{1}{P_{\rm in}^{\rm K}} \int\limits_0^{2\pi} \left(\dot{X}_\alpha\right)_{\rm 1pN} \left(\frac{dt}{df}\right)_{\rm K} \left(\frac{dt}{dF}\right)_{\rm K} df\ dF,
\end{equation}
where
\begin{flalign}
    P_{\rm out}^{\rm quad} = & \int\limits_0^{2\pi} \left \lbrack \left(\frac{dt}{dF}\right)_{\rm quad} + W_\beta^{\frac{7}{2}0}\frac{\partial}{\partial \tilde{X}_\beta} \left(\frac{dt}{dF}\right)_{\rm K} \right \rbrack dF \approx \int\limits_0^{2\pi} \left \langle \left(\frac{dt}{dF}\right)_{\rm quad} + W_\beta^{\frac{7}{2}0}\frac{\partial}{\partial \tilde{X}_\beta} \left(\frac{dt}{dF}\right)_{\rm K} \right \rangle_{\rm in} dF && \\\nonumber
    = & \int\limits_0^{2\pi} \frac{1}{P_{\rm in}} \int\limits_0^{2\pi} \left \lbrack \left(\frac{dt}{dF}\right)_{\rm quad} + W_\beta^{\frac{7}{2}0}\frac{\partial}{\partial \tilde{X}_\beta} \left(\frac{dt}{dF}\right)_{\rm K} \right \rbrack \left(\frac{dt}{df}\right)\ df\ dF && \\\nonumber
    = & P_{\rm out}^{\rm K} \left( \frac{p_1^2}{p_2^2} \right) \left(\frac{\eta}{32 \ell_2^2 \ell_1^4 (1+\ell_2)^2}\right) \bigg(-3 (-8 (A_1-A_4) (A_1+A_4) \ell _1^2 \ell _2^5-3 \ell _2^4 (3 (A_1^2+3 A_4^2+4) \ell _1^2-20)-6 \ell _2^3 (3 && \\\nonumber
    & \times (A_1^2+3 A_4^2+4) \ell _1^2-20)+4 \ell _2^2 (3 (A_1^2+A_4^2+2) \ell _1^2-10)+6 (7 A_1^2+13 A_4^2+20) \ell _1^2 \ell _2+3 (7 A_1^2+13 A_4^2+20) \ell _1^2 && \\\nonumber
    & -A_3^2 (4 \ell _1^2-5) (\ell _2 (\ell _2 (\ell _2 (\ell _2 (8 \ell _2-27)-54)+12)+78)+39)+A_2^2 (4 \ell _1^2-5) (\ell _2 (\ell _2 (\ell _2 (\ell _2 (8 \ell _2+9)+18) && \\\nonumber
    & -12)-42)-21)-100 (2 \ell _2+1))-6 \ell _2^5 (-5 e_1^2 (\cos (2 \iota )+3) \cos (2 \omega _1) \cos (2 \omega _2)-20 e_1^2 \cos (\iota ) \sin (2 \omega _1) \sin (2 \omega _2)&&\\\nonumber
    & +2 (3 \ell _1^2-5) \sin ^2(\iota ) \cos (2 \omega _2))\bigg),
\end{flalign}
where
\begin{align}\label{eq:constantsapp}
\begin{split}
    A_1 & =  \cos\iota \cos\omega_1 \cos\omega_2+\sin\omega_1\sin\omega_2 \\
    A_2 & =  \cos\iota \cos\omega_2\sin\omega_1 - \cos\omega_1\sin\omega_2 \\
    A_3 & =  \cos\omega_1 \cos\omega_2 + \cos\iota \sin\omega_1\sin\omega_2 \\
    A_4 & =  \cos\omega_2\sin\omega_1 - \cos\iota \cos\omega_1\sin\omega_2.
\end{split}
\end{align}
These cross terms are equal to the usual secular quadrupole terms times a multiplicative factor:
\begin{equation}
    \frac{d \tilde{X}_\alpha}{dt} = -\frac{P_{\rm out}^{\rm quad}}{P_{\rm in}^{\rm K}} \left(\frac{d \tilde{X}_\alpha}{dt}\right)_{\rm quad}
\end{equation}
\end{subsubsection}
\end{subsection}

\begin{subsection}{Indirect cross terms due to periodic 1pN perturbations} \label{sec:IDpNB}
These cross terms come from average-free, periodic 1pN perturbations which combine with perturbations from $\bm{a}_{\rm quad}$. The inner binary periodic perturbations do not generate secular effects,
\begin{equation}
    \frac{d\tilde{X}_\alpha}{dt} = \frac{1}{P_{\rm out}^{\rm K}} \int \limits_0^{2\pi}  \sum\limits_{\beta=1}^5 W_\beta^{01}\  \frac{\partial( Q_\alpha^{(0)} )_{\rm quad}}{\partial \tilde{X}_\beta}\ dF = 0
\end{equation}
The outer binary periodic perturbations generate secular effects which read
\begin{equation}
    \frac{d\tilde{X}_\alpha}{dt} = \frac{1}{P_{\rm out}^{\rm K}} \int \limits_0^{2\pi} \sum\limits_{\beta=6}^{10} W_\beta^{\frac{5}{2}1}\  \frac{\partial( Q_\alpha^{(0)} )_{\rm quad}}{\partial \tilde{X}_\beta}\ dF.
\end{equation}

\begin{subsubsection}{Cross terms from periodic 1pN effects on the outer binary}

\begin{flalign}\label{eq:dp1dtIDpNBa}
\frac{dp_1}{dt} = & \frac{15 e_1^2 G^{3/2} m_3 p_1^{5/2} \ell _2^3}{32 c^2 \sqrt{m} M p_2^4 \ell _1^4} \bigg(\sin (\omega _1) \cos (\omega _1) (\cos ^2(\omega _2) (L_{\ell _2}-\cos ^2(\iota ) K_{\ell _2})+\sin ^2(\omega _2) (K_{\ell _2}-\cos ^2(\iota ) L_{\ell _2})) && \\\nonumber
& -2 F_{\ell _2} \cos (\iota ) \sin (2 \omega _2) \cos (2 \omega _1)\bigg)
\end{flalign}

\begin{flalign}\label{eq:de1dtIDpNBa}
\frac{de_1}{dt} = & \frac{15 e_1 m_3 \ell _2^3 (G p_1){}^{3/2}}{256 c^2 M p_2^4 \ell _1^2 \sqrt{m}} \bigg(8 F_{\ell _2} \cos (\iota ) \sin (2 \omega _2) \cos (2 \omega _1)+\sin (\omega _1) \cos (\omega _1) ((\cos (2 \iota )+3) \cos (2 \omega _2) (K_{\ell _2}-L_{\ell _2}) && \\\nonumber
& -2 \sin ^2(\iota ) (K_{\ell _2}+L_{\ell _2}))\bigg)
\end{flalign}

\begin{flalign}\label{eq:di1dtIDpNBa}
\frac{d\iota_1}{dt} = & \frac{3 m_3 \ell _2^3 (G p_1){}^{3/2} \sin (\iota ) }{64 c^2 \sqrt{m} M p_2^4 \ell _1^4} \bigg(F_{\ell _2} \sin (2 \omega _2) (5 e_1^2 \cos (2 \omega _1)-3 \ell _1^2+5)+\frac{5}{4} e_1^2 \cos (\iota ) \sin (2 \omega _1) (\cos (2 \omega _2) (K_{\ell _2}-L_{\ell _2}) && \\\nonumber
& +K_{\ell _2}+L_{\ell _2})\bigg)
\end{flalign}

\begin{flalign}\label{eq:domega1dtIDpNBa}
\frac{d\omega_1}{dt} = & -\frac{3 m_3 \ell _2^3 (G p_1){}^{3/2}}{64 c^2 \sqrt{m} M p_2^4 \ell _1^4} \bigg(20 F_{\ell _2} \sin (\omega _2) \sin (\omega _1) \cos (\omega _1) \cos (\omega _2) (e_1^2 \sin (\iota ) \cot (\iota _1)+2 \ell _1^2 \cos (\iota ))+2 \ell _1^2 H_{\ell _2} && \\\nonumber
& +\sin ^2(\omega _1) (\sin ^2(\omega _2) ((5-4 \ell _1^2) \sin (\iota ) \cos (\iota ) \cot (\iota _1) L_{\ell _2}-\ell _1^2 (K_{\ell _2}-4 \cos ^2(\iota ) L_{\ell _2}))+\cos ^2(\omega _2) ((5-4 \ell _1^2) && \\\nonumber
& \times \sin (\iota ) \cos (\iota ) \cot (\iota _1) K_{\ell _2}-\ell _1^2 (L_{\ell _2}-4 \cos ^2(\iota ) K_{\ell _2})))+\ell _1^2 \cos ^2(\omega _1) (\cos ^2(\omega _2) (\sin (\iota -\iota _1) \cos (\iota ) \csc (\iota _1) K_{\ell _2} && \\\nonumber
& +4 L_{\ell _2})+\sin ^2(\omega _2) (4 K_{\ell _2}+\sin (\iota -\iota _1) \cos (\iota ) \csc (\iota _1) L_{\ell _2}))\bigg)
\end{flalign}

\begin{flalign}\label{eq:dOmega1dtIDpNBa}
\frac{d\Omega_1}{dt} = & \frac{3 m_3 \ell _2^3 (G p_1){}^{3/2} \sin (\iota ) \csc (\iota _1)}{64 c^2 \sqrt{m} M p_2^4 \ell _1^4}  \bigg(5 e_1^2 F_{\ell _2} \sin (2 \omega _1) \sin (2 \omega _2)+\frac{1}{4} \cos (\iota ) (-5 e_1^2 \cos (2 \omega _1)-3 \ell _1^2+5) (\cos (2 \omega _2) && \\\nonumber
& \times (K_{\ell _2}-L_{\ell _2})+K_{\ell _2}+L_{\ell _2})\bigg),
\end{flalign}
where
\begin{align}
    F_{\ell_2} & \equiv M^2(-32+6\ell_2^2) + m m_3 (29-11\ell_2^2) \\
    H_{\ell_2} & \equiv 8 M^2(2-5\ell_2^2) + m m_3(7+\ell_2^2) \\
    K_{\ell_2} & \equiv 4M^2(12+7\ell_2^2)+m m_3(-65+21\ell_2^2) \\
    L_{\ell_2} & \equiv M^2(-80+52\ell_2^2)+m m_3(51-23\ell_2^2).
\end{align}

\end{subsubsection}
\end{subsection}

\begin{subsection}{Indirect cross terms due to periodic quadrupole perturbations} \label{sec:IDNQ}
These cross terms come from average-free, periodic quadrupole perturbations which combine with perturbations from $\bm{a}_{\rm 1pN}$. The secular effects from periodic perturbations on the inner binary are
\begin{equation}
    \frac{d\tilde{X}_\alpha}{dt} = \frac{1}{P_{\rm out}^{\rm K}} \int \limits_0^{2\pi}\sum\limits_{\beta=1}^5 W_\beta^{30}\  \frac{\partial( Q_\alpha^{(0)} )_{\rm 1pN}}{\partial \tilde{X}_\beta}\ dF,
\end{equation}
and on the outer binary are
\begin{equation}
    \frac{d\tilde{X}_\alpha}{dt} = \frac{1}{P_{\rm out}^{\rm K}} \int \limits_0^{2\pi} \sum\limits_{\beta=6}^{10} W_\beta^{\frac{7}{2}0}\  \frac{\partial( Q_\alpha^{(0)} )_{\rm 1pN}}{\partial \tilde{X}_\beta}\ dF.
\end{equation}
We find that only $\omega_1$ is affected, with no secular effects on the other elements.
\begin{subsubsection}{Cross terms from periodic quadrupole effects on the outer binary}
\begin{flalign}
\frac{d\omega_1}{dt} = & \frac{9 \eta  G^{3/2} m^{3/2}}{32 c^2 e_2^4 \sqrt{p_1} p_2^2 \ell _1 \ell _2^2} \bigg(\ell _2^6 (\ell _1^2 (7 A_1^2-43 A_4^2-36)+60)+\ell _2^4 (3 \ell _1^2 (13 A_1^2+31 A_4^2+44)-220)-3 \ell _1^2 \ell _2^2 (17 A_1^2 && \\\nonumber
& +35 A_4^2+52)+3 \ell _1^2 (7 A_1^2+13 A_4^2+20)-8 \ell _1^2 \ell _2^7 (A_1-A_4) (A_1+A_4)-8 \ell _1^2 \ell _2^5 (A_1-A_4) (A_1+A_4) && \\\nonumber
& +A_2^2 (4 \ell _1^2-5) (\ell _2-1){}^2 (\ell _2 (\ell _2 (\ell _2 (\ell _2 (8 \ell _2+9)+18)-12)-42)-21)-A_3^2 (4 \ell _1^2-5) (\ell _2-1){}^2 && \\\nonumber
& \times (\ell _2 (\ell _2 (\ell _2 (\ell _2 (8 \ell _2-27)-54)+12)+78)+39)+20 (13 \ell _2^2-5)\bigg),
\end{flalign}
where $A_1$, $A_2$, $A_3$, and $A_4$ are defined in Eq.~(\ref{eq:constantsapp}).
\end{subsubsection}

\begin{subsubsection}{Cross terms from periodic quadrupole effects on the inner binary}
\begin{flalign}
\frac{d\omega_1}{dt} = \frac{15 G^{3/2} m m_3 (1-\ell_2)(1+2\ell_2) e_1^2 C_1}{4 c^2 \sqrt{M} p_1 p_2^{3/2} \ell_1 (1+\ell_2)},
\end{flalign}
where $C_1 =  A_1 A_3 - A_2 A_4$.
\end{subsubsection}

\end{subsection}
\end{widetext}

\end{document}